\documentclass[twocolumn]{aastex701}
\usepackage{enumitem}
\usepackage{multirow}
\usepackage{comment}
\usepackage{tabularray}

\shorttitle{GOATS: Gemini Observation and Analysis of Targets System}
\shortauthors{Soraisam at al.}
\begin{document}

\title{GOATS: The next generation software infrastructure for time-domain astronomy at Gemini/NOIRLab\\
{\small Application to alerts from Vera C.~Rubin Observatory's Legacy Survey of Space and Time}
}

\correspondingauthor{Monika Soraisam}
\author[0000-0001-6360-992X]{Monika D.\ Soraisam}
\affiliation{International Gemini Observatory / NSF NOIRLab, 670 N.\ A’ohoku Place, Hilo, Hawai’i, 96720, USA}
\email[show]{monika.soraisam@noirlab.edu}

\author[0009-0006-3587-3380]{Louis Avner}
\affiliation{NSF NOIRLab, 950 N.\ Cherry Ave.\, Tucson, AZ 85719, USA}
\email{}

\author[0009-0007-1807-1340]{Miguel G\'{o}mez}
\affiliation{NSF NOIRLab, Recinto de AURA Avda Juan Cisternas 1500, La Serena, Chile}
\email{}

\author[0000-0002-9123-0068]{William D. Vacca}
\affiliation{International Gemini Observatory / NSF NOIRLab, 950 N.\ Cherry Ave.\, Tucson, AZ 85719, USA}
\email{}

\author[0000-0002-5665-376X]{Bryan W. Miller}
\affiliation{International Gemini Observatory / NSF NOIRLab, Casilla 603, La Serena, Chile}
\email{}

\author[0000-0002-4434-2307]{Andrew Stephens}
\affiliation{International Gemini Observatory / NSF NOIRLab, 670 N.\ A’ohoku Place, Hilo, Hawai’i, 96720, USA}
\email{}

\author[]{Arturo N\'{u}\~{n}ez}
\affiliation{NSF NOIRLab, Recinto de AURA Avda Juan Cisternas 1500, La Serena, Chile}
\email{}

%%%%%  %%%%%%
\author[0000-0003-1120-5178]{Andrew Adamson}
\affiliation{International Gemini Observatory / NSF NOIRLab, 670 N.\ A’ohoku Place, Hilo, Hawai’i, 96720, USA}
\email{}

\author[0000-0001-7124-4094]{C\'{e}sar Brice\~{n}o}
\affiliation{SOAR Telescope / NSF NOIRLab, Casilla 603, La Serena, Chile}
\email{}

\author[]{Hern\'{a}n Chacana}
\affiliation{NSF NOIRLab, Recinto de AURA Avda Juan Cisternas 1500, La Serena, Chile}
\email{}

\author[0000-0002-2651-7038]{Guillermo Damke}
\affiliation{Cerro Tololo Inter-American Observatory / NSF NOIRLab, Casilla 603, La Serena, Chile}
\email{}

\author[0009-0002-2806-9379]{Nicol\'{a}s Esquivel}
\affiliation{NSF NOIRLab, Recinto de AURA Avda Juan Cisternas 1500, La Serena, Chile}
\email{}

\author[0009-0007-4329-4765]{Paul Hirst}
\affiliation{International Gemini Observatory / NSF NOIRLab, 670 N.\ A’ohoku Place, Hilo, Hawai’i, 96720, USA}
\email{}

\author[0000-0002-6633-7891]{Kathleen Labrie}
\affiliation{International Gemini Observatory / NSF NOIRLab, 670 N.\ A’ohoku Place, Hilo, Hawai’i, 96720, USA}
\email{}

\author[0000-0001-6685-0479]{Thomas Matheson}
\affiliation{NSF NOIRLab, 950 N.\ Cherry Ave.\, Tucson, AZ 85719, USA}
\email{}

\author[]{Chadd Myers}
\affiliation{NSF NOIRLab, 950 N.\ Cherry Ave.\, Tucson, AZ 85719, USA}
\email{}

\author[0000-0002-7052-6900]{Robert Nikutta}
\affiliation{NSF NOIRLab, 950 N.\ Cherry Ave.\, Tucson, AZ 85719, USA}
\email{}

\author[0000-0002-6839-4881]{Abhijit Saha}
\affiliation{NSF NOIRLab, 950 N.\ Cherry Ave.\, Tucson, AZ 85719, USA}
\email{}

\author[0000-0001-8589-4055]{Chris Simpson}
\affiliation{International Gemini Observatory / NSF NOIRLab, 670 N.\ A’ohoku Place, Hilo, Hawai’i, 96720, USA}
\email{}

\author[0009-0001-0560-2954]{Olesja Smirnova}
\affiliation{International Gemini Observatory / NSF NOIRLab, Casilla 603, La Serena, Chile}
\email{}

\author[0000-0002-1912-3057]{D.\ J.\ Teal}
\affiliation{NSF NOIRLab, 950 N.\ Cherry Ave.\, Tucson, AZ 85719, USA}
\email{}

\author[0009-0005-0192-3784]{Sergio Troncoso}
\affiliation{NSF NOIRLab, Recinto de AURA Avda Juan Cisternas 1500, La Serena, Chile}
\email{}

\author[]{James Turner}
\affiliation{International Gemini Observatory / NSF NOIRLab, Casilla 603, La Serena, Chile}
\email{}

\author[]{Sebasti\'{a}n Vicencio}
\affiliation{NSF NOIRLab, Recinto de AURA Avda Juan Cisternas 1500, La Serena, Chile}
\email{}

%%%%%  %%%%%%
\author[]{Hubert Condoretti}
\affiliation{NSF NOIRLab, 950 N.\ Cherry Ave.\, Tucson, AZ 85719, USA}
\email{}

\author[0000-0002-2968-2418]{Scott Dahm}
\affiliation{International Gemini Observatory / NSF NOIRLab, 670 N.\ A’ohoku Place, Hilo, Hawai’i, 96720, USA}
\email{}

%%%%%%%%  %%%%%%%
\author[0000-0002-2184-6430]{Tom\'{a}s Ahumada}
\affiliation{SOAR Telescope / NSF NOIRLab, Casilla 603, La Serena, Chile}
\email{}

\author[0000-0001-8018-5348]{Eric Bellm}
\affiliation{DIRAC Institute, Department of Astronomy, University of Washington, Seattle, WA 98195, USA}
\email{}

\author[0000-0002-8622-4237]{Robert Blum}
\affiliation{NSF-DOE Vera C.\ Rubin Observatory / NSF NOIRLab, 950 N.\ Cherry Ave.\, Tucson, AZ 85719, USA}
\email{}

\author[0000-0001-7101-9831]{Aleksandar Cikota}
\affiliation{International Gemini Observatory / NSF NOIRLab, Casilla 603, La Serena, Chile}
\email{}

\author[]{Hannah Crayton}
\affiliation{NSF NOIRLab, Casilla 603, La Serena, Chile}
\email{}

\author[]{Diego G\'{o}mez}
\affiliation{NSF NOIRLab, Casilla 603, La Serena, Chile}
\email{}

\author[0000-0002-9154-3136]{Melissa L.~Graham}
\affiliation{University of Washington, Department of Astronomy, Box 351580, Seattle, WA 98195, USA}
\affiliation{Institute for Data-intensive Research in Astrophysics and Cosmology, University of Washington, 3910 15th Avenue NE, Seattle, WA 98195, USA}
\email{}

\author[0000-0002-0856-3663]{Stephen Heathcote}
\affiliation{Cerro Tololo Inter-American Observatory / NSF NOIRLab, Casilla 603, La Serena, Chile}
\email{}

\author[0000-0003-0800-8755]{Leanne P.\ Guy}
\affiliation{NSF-DOE Vera C.\ Rubin Observatory / NSF NOIRLab, Casilla 603, La Serena, Chile}
\email{}

\author[0000-0002-9667-2244]{Fredrik Rantakyr\"{o}}
\affiliation{International Gemini Observatory / NSF NOIRLab, Casilla 603, La Serena, Chile}
\email{}

\author[0000-0002-2234-749X]{Kevin Reil}
\affiliation{SLAC National Accelerator Laboratory, 2575 Sand Hill Rd., Menlo Park, CA 94025, USA}
\email{}

\author[0000-0001-6279-0552]{Rachel Street}
\affiliation{Las Cumbres Observatory, 6740 Cortona Drive, Suite 102, Goleta, CA 93117, USA}
\email{}

\author[0000-0002-9121-3436]{Sandrine Thomas}
\affiliation{NSF-DOE Vera C.\ Rubin Observatory / NSF NOIRLab, 950 N.\ Cherry Ave.\, Tucson, AZ 85719, USA}
\email{}

\author[0000-0002-2726-6971]{Sim\'{o}n Torres}
\affiliation{NSF NOIRLab, Casilla 603, La Serena, Chile}
\email{}

\author[0000-0003-4341-6172]{A.\ Katherina Vivas}
\affiliation{Cerro Tololo Inter-American Observatory / NSF NOIRLab, Casilla 603, La Serena, Chile}
\email{}

\begin{abstract}
Time-domain and multimessenger astronomy (MMA/TDA) targets demand rapid-response follow-up observations. In many cases, it is the only way to make discoveries and advance our understanding of the astrophysical phenomena, for example, kilonovae accompanying gravitational waves from compact object mergers, shock breakout in supernovae, prompt emission from GRBs, etc. Presently the MMA/TDA follow-up workflow requires wrangling disparate software packages and user interfaces. We present an end-to-end software tool for the community, the Gemini Observation and Analysis of Targets System (GOATS), which unifies and simplifies the workflow, particularly for Gemini follow-up observations. GOATS achieves this by integrating services from Gemini Observatory and its parent organization, NSF NOIRLab. From a single platform, GOATS enables enhanced target selection via NOIRLab's ANTARES alert broker, triggering of Gemini (and other facilities within the Astronomical Event Observatory Network), automated data retrieval from the Gemini Observatory Archive, and interactive data reduction and analysis through Gemini's DRAGONS software and NOIRLab's Astro Data Lab science platform. 
GOATS was successfully deployed in an end-to-end demonstration of real-time follow-up of Rubin/LSST alerts with NOIRLab facilities. As part of this demonstration, we selected targets from the Rubin alert stream and triggered follow-up observations within minutes of the Rubin detections. We obtained spectra for several targets and classified them as supernova of various types (Ia, IIP, Ib/c) with redshifts ranging from 0.05 to 0.35. By eliminating the need to manually connect tools and automating repetitive tasks, GOATS lowers the entry barrier and allows users to focus on the scientific interpretation of the observation results. 

\end{abstract}

\keywords{Astronomy software --- Astronomy data analysis --- Time domain astronomy --- Sky surveys}

\section{Introduction}\label{sec:intro}
Time-domain astronomy (TDA) has made rapid progress in the past two decades with various all-sky surveys routinely discovering variable and transient sources, throwing light on diverse astrophysical phenomena ranging from cosmology to stellar physics, exoplanets, and solar system objects. The field is poised to advance further with current and upcoming surveys, such as ATLAS \citep{Tonry-2018}, ZTF \citep{Bellm-2019}, Young Supernova Experiment \citep{Jones-2021}, BlackGEM \citep{Groot-2024}, Large Area Survey Telescope \citep{Ofek-2023}, Argus Array \citep{Law-2022}, CHIME \citep{Amiri-2022}, Deep Synoptic Array \citep{DSA-2024} and even space-based surveys with, e.g., Roman \citep{Spergel-2015}, and ULTRASAT \citep{Ultrasat}. Capping off this list is Rubin Observatory's Legacy Survey of Space and Time (LSST; \citealt{LSST}), which is slated to revolutionize the field of TDA during its 10~years of operation, reaching deeper than most time-domain surveys. All of these facilities promise a rich yield of discoveries -- supernovae, GRBs, tidal disruption events, AGN flares, fast radio bursts, kilonovae, novae, variable stars, moving objects, etc.\ and also unknown phenomena. 

More recently, the first joint discovery of a binary neutron star merger (GW170817) in gravitational waves and electromagnetic radiation \citep{Abbott-2017} ushered in a new era of multi-messenger astronomy (MMA) dedicated to exploring the universe using multiple cosmic messengers---gravitational waves, photons, neutrinos, and cosmic rays. Nevertheless, the search for and characterization of the electromagnetic counterpart of the MMA events is essentially a TDA pursuit. In fact, NOIRLab facilities, such as the Gemini and SOAR Observatories played a pivotal role in the follow-up of the kilonova counterpart of GW170817; both GMOS and FLAMINGOS-2 instruments on Gemini-South, and the Goodman instrument on SOAR, were used for imaging and spectroscopic follow-up of the counterpart \citep{Blanchard2017} and infrared spectroscopic data obtained with FLAMINGOS-2 established that heavy elements were synthesized in this merger \citep{Chornock-2017}. 

\begin{figure*}[h]
    \centering
    \includegraphics[width=0.8\textwidth]{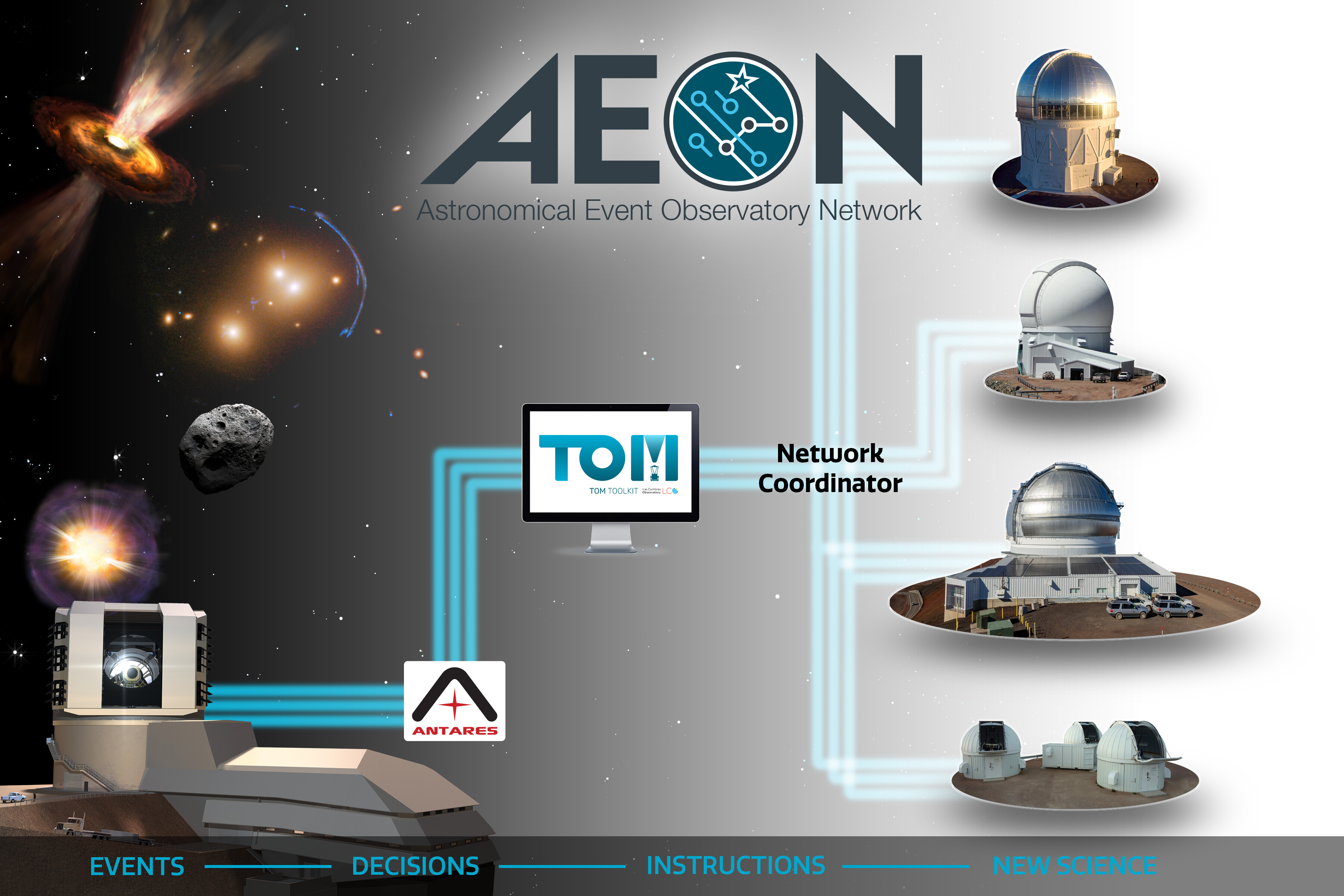}
    \caption{Schematic of the Astronomical Event Observatory Network (AEON), which is a collection of world-class facilities, including those of NOIRLab (a founding partner of AEON), for efficient follow-up of MMA/TDA events. 
    Image credit: NOIRLab/NSF/AURA/P.\,Marenfeld.}
    \label{fig:aeon}
\end{figure*}

With planned upgrades to existing neutrino facilities and networks of gravitational wave detectors and considerations for more powerful facilities in the next decade \citep{Corsi-2024, LISA, Aartsen-2021}, the importance of MMA/TDA will only increase. This is clearly evidenced by the Decadal Survey on Astronomy and Astrophysics 2020 \citep{Astro2020} identifying MMA/TDA as one of the science priorities. The groundbreaking scientific yield promised by MMA/TDA will, however, materialize only if the follow-up tools and resources required for characterizing the sources are in place.

\subsection{Technical motivation}\label{sec:sci}

Follow-up of MMA/TDA events is fraught with unique technical challenges, requiring a \emph{network} of facilities with diverse capabilities, e.g., field of view, observation depth and mode (imaging vs.\ spectroscopy), wavelength coverage, sky coverage, etc. (see, e.g., \citealt{WOU_WP}). Software is the key to realizing this network and enabling coordination between the facilities. These MMA/TDA events are often rapidly evolving and therefore fast response of every component of the follow-up infrastructure is essential. Even if there is a reliable observing facility able to swap instruments and get on the target within minutes, the community will not be able to benefit from this capability without equally agile software tools to interface with the facility. Importantly, this interface must span the whole life cycle of the experiment: a user should be able to promptly communicate the target information to the facility, and the latter should promptly communicate the status of the observation and availability of data to the user.

Various state-of-the-art services and tools have been developed by NOIRLab and Gemini to facilitate the steps of the MMA/TDA follow-up workflow. NOIRLab, in partnership with Las Cumbres Observatory (Las Cumbres), founded the Astronomical Event Observatory Network (AEON; Fig.~\ref{fig:aeon}), which serves as a collection of world-class telescopes accessible on demand via an application programming interface (API) for rapid follow-up of MMA/TDA alerts. The Gemini 8-m, SOAR 4-m, and Blanco 4-m telescopes are all currently AEON-enabled such that users can submit follow-up observations through an API to NOIRLab observing databases. However, this API provides only limited support (see Section~\ref{sec:goats}). 

\begin{figure*}
    \centering
    \includegraphics[width=0.8\linewidth]{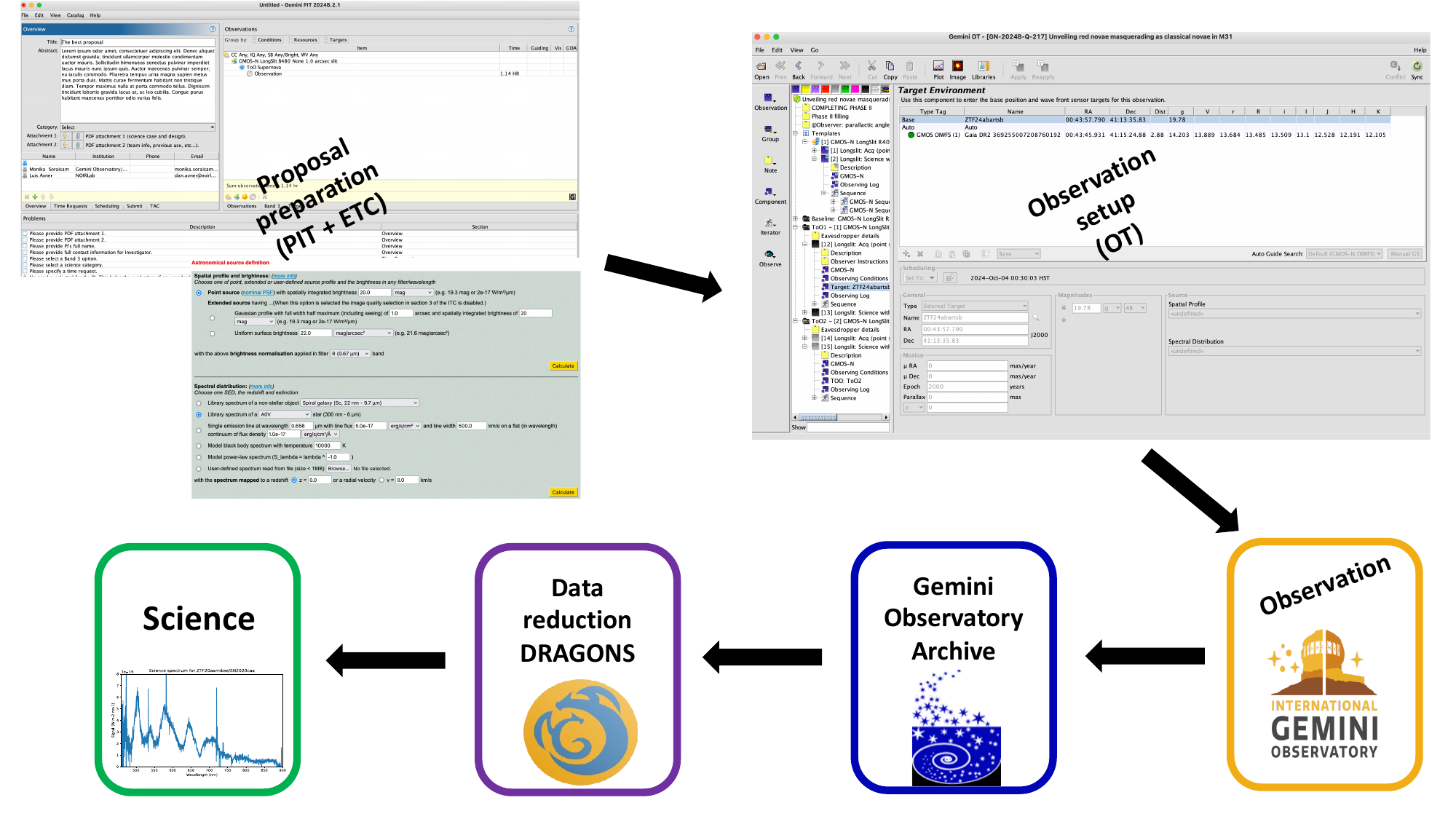}
    \caption{Various interfaces that a user needs to interact with in order to plan, gather and reduce Gemini observation data.}
    \label{fig:phases123}
\end{figure*}

Other pieces of infrastructure developed and maintained in-house that are critical to the MMA/TDA enterprise include: the ANTARES event broker \citep{antares}, which winnows down the barrage of alerts from MMA/TDA surveys; data archives, e.g., Gemini Observatory Archive (GOA; \citealt{GOA}), Astro Data Archive \citep{McManus}; the Astro Data Lab science platform \citep{DataLab2014, DataLab2020} and the Rubin Science Platform \citep{OMullane-2024}, where archival big-survey data and computing resources are co-located; and data reduction software, such as DRAGONS for Gemini \citep{DRAGONS}, the Goodman Spectroscopic Pipeline for SOAR \citep[GSP;][]{Torres}, and the Blanco-DECam Community Data Pipeline \citep{community_pipeline}, which process the follow-up data to make it science-ready. However, a lack of integration between these tools has posed a major challenge in the MMA/TDA infrastructure of NOIRLab/Gemini, limiting the scientific return. 

As an example, Fig.~\ref{fig:phases123} demonstrates the different software and interfaces that a user must install, access online, and learn to operate in order to execute a Gemini observation and perform scientific analysis. First, one downloads a Java-based tool, the Phase I Tool (PIT), to use for preparing and submitting the Gemini proposal. During proposal preparation, the web-based integration time calculator is used to estimate the time requests. After submission, the proposal is reviewed by the time allocation committee. If awarded time, it goes through {\it Phase II}, where the user prepares the observations before the start of the semester, and Gemini staff review them to ensure the observing details are complete and consistent with the proposal. The {\it Phase II} process requires a second Java-based application, the Observing Tool (OT), which the user must again install. The verified observations then enter the queue for scheduling during the semester. 

For MMA/TDA programs, where the targets are not known in advance, the user must return to the OT once targets are identified to update the prepared observation with target information and trigger for rapid scheduling; this mode of observing is known as Target of Opportunity (ToO). After the observation completes, the data are ingested into GOA---typically within a minute---and the user must log in to download the files. To reduce these data, the user must have DRAGONS installed before finally reaching the stage of scientific analysis. For a scientific field that depends on rapid follow-up, such a workflow is a significant barrier, particularly for new users. 

\begin{figure*}
    \centering
    \includegraphics[width=140mm]{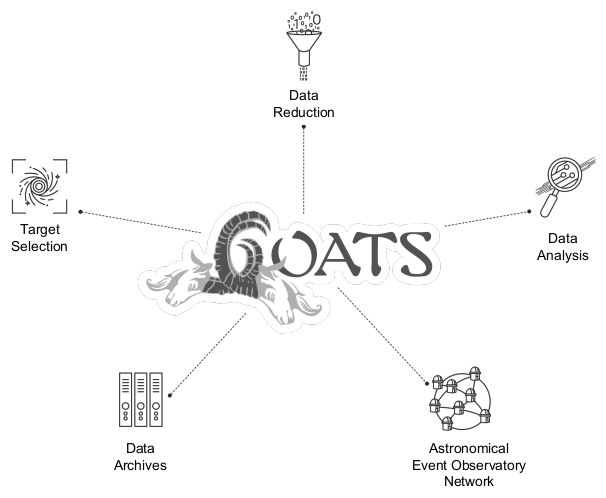}
    \caption{Time-domain and multi-messenger astronomy follow-up services integrated by GOATS.}
    \label{fig:goats_general}
\end{figure*}

\begin{figure*}
    \centering
    \includegraphics[width=150mm]{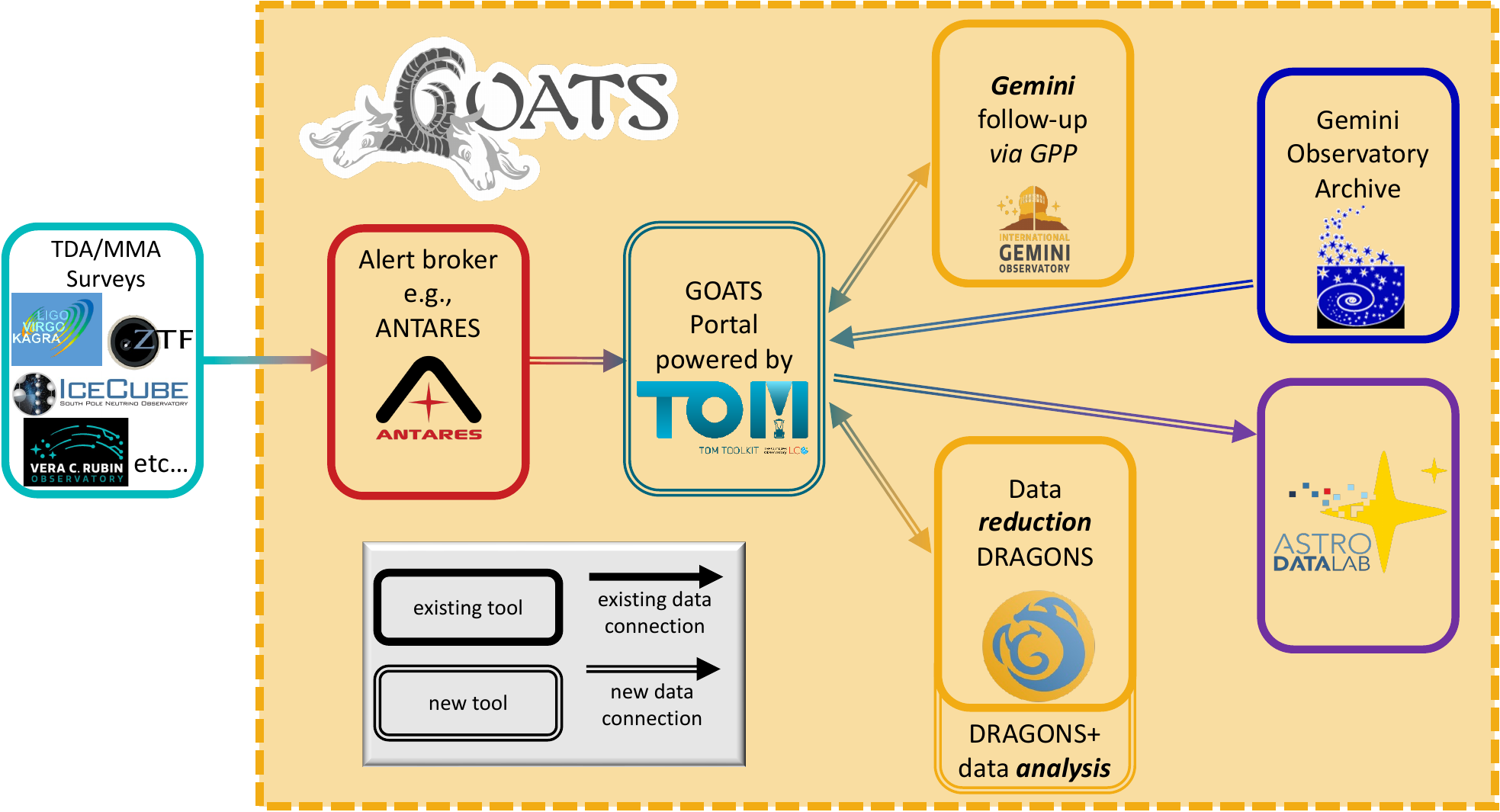}
    \caption{
    Schematic of GOATS, showing both existing and newly developed tools and data connections supporting follow-up observations with Gemini, thereby increasing efficiency and lowering the entry barrier. Users can send target information directly from the ANTARES broker portal, trigger Gemini observations, retrieve data from the Gemini Observatory Archive, and perform interactive data reduction with DRAGONS and subsequent analysis---all within the GOATS platform.
    }
    \label{fig:goats_schema}
\end{figure*}

\subsection{The Gemini Observation and Analyis of Targets System (GOATS): Key deliverables}\label{sec:deliverables}
A way to solve the above coordination problem is to provide a user interface that glues together the disparate services. The Las Cumbres TOM-Toolkit \citep{TOMToolkit} has been an important contribution to this end, providing a framework for building web applications, so called Target and Observation Managers (TOMs), to manage targets. Through software plugins, including those contributed by NOIRLab, users can spin up a TOM to select targets from community alert brokers and trigger follow-up observations on the AEON facilities (e.g., Las Cumbres, Gemini, and SOAR). Various teams have built TOM systems customized specifically for their respective collaborations, such as SkyPortal \citep{Coughlin-2023}, SAGUARO~TOM \citep{Hosseinzadeh-2024}, Black Hole TOM \citep{Wyrzykowski-2024}, and Supernova Exchange \citep{Pellegrino-2024}; the latter three are built upon the TOM-Toolkit library. However, the existing TOM-Toolkit provides relatively limited functionality for managing follow-up data and requires users or admins to manually configure settings files with facility-specific URLs, API keys, and other parameters. With the base library, users can only download data from the Las Cumbres archive or import files locally for display.  Moreover, there is no support for data reduction, which poses a critical gap in fully closing the follow-up loop. 

With the aim to unify and simplify the Gemini follow-up observation workflow for the broader community, we have developed the Gemini Observation and Analysis of Targets System (GOATS). Built using the TOM-Toolkit library, GOATS integrates the MMA/TDA services of Gemini Observatory and the larger NOIRLab organization into a single platform that astronomers can use out-of-the-box. Beyond the facilities already supported by the base TOM-Toolkit library, GOATS provides end-to-end automation for Gemini programs---from target selection to triggering of follow-up observation, data retrieval, and data reduction and analysis (Fig.~\ref{fig:goats_general}). It thus serves as a one-stop shop for the follow-up needs of the MMA/TDA community while lowering the entry barrier for adopting the existing tools and services of Gemini/NOIRLab. Moreover, the system will also be valuable to a wide variety of non-MMA/TDA science cases that would benefit from automation.

The key deliverables of GOATS are the following.
\begin{enumerate}[label=(\roman*)]
    \item Integrate the ANTARES portal for direct target selection (Section~\ref{sec:antares}).
    \item Support triggering of Gemini observations with all the flexibility for defining and updating the observation configuration as well as receiving observation status (Section~\ref{sec:gpp}). GOATS can also be used to trigger other AEON facilities (Section~\ref{sec:aeon}).
    \item Integrate the Gemini Observatory Archive to automatically retrieve data, including proprietary data (Section~\ref{sec:goa}).
    \item Integrate the DRAGONS data reduction software for Gemini (Section~\ref{sec:dragons}).
    \item Interface with the NOIRLab Astro Data Lab science platform (Section~\ref{sec:datalab}). 
    \item Integrate data analysis tools, particularly for spectroscopic data (Section~\ref{sec:analysis}).
\end{enumerate}

Fig.~\ref{fig:goats_schema} shows the building blocks of GOATS and the flow of data and information within the platform.

\section{GOATS software development}\label{sec:goats}
\begin{figure}
    \centering
    \includegraphics[width=\linewidth]{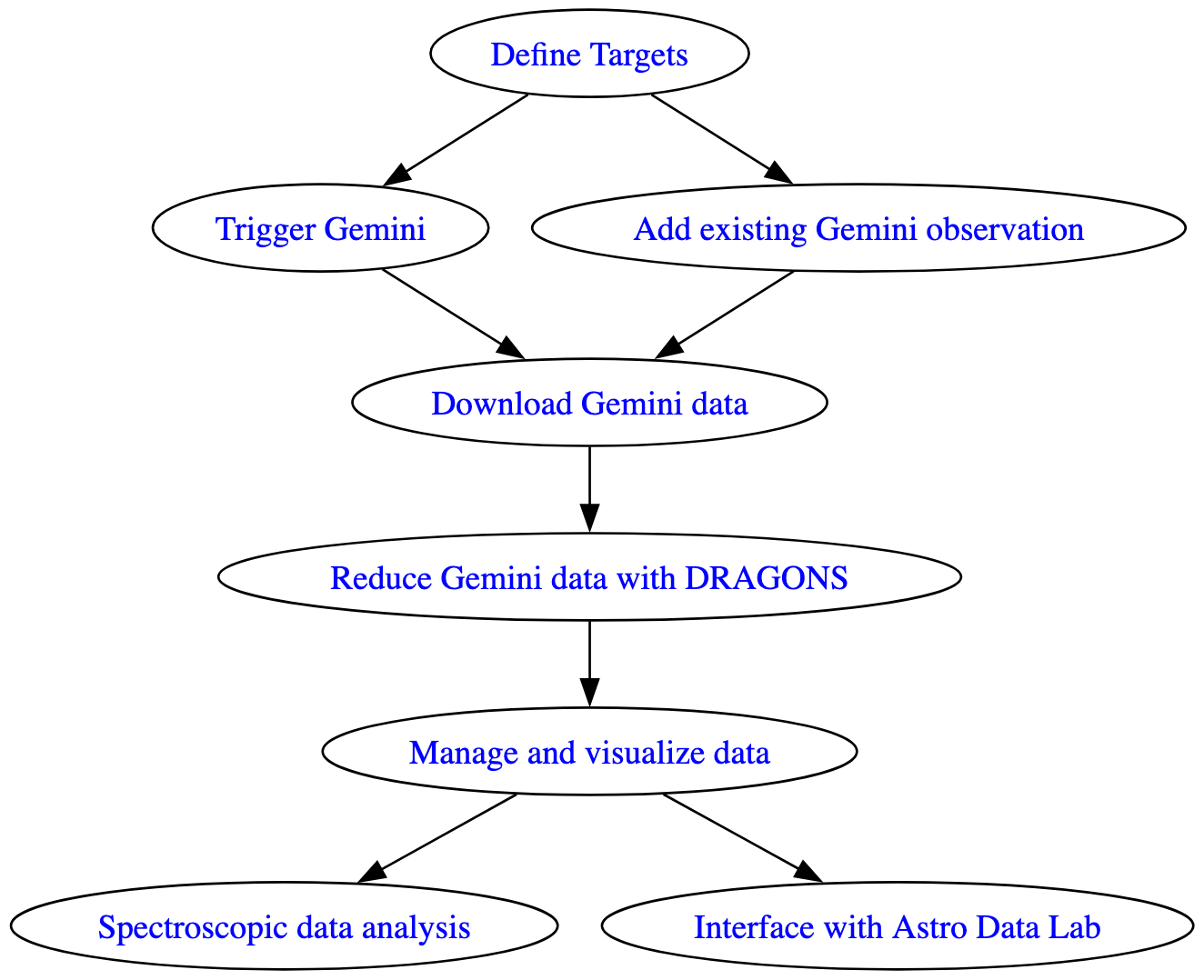}
    \caption{Flowchart showing the typical workflow in GOATS for Gemini observations. The workflow is similar for other supported AEON facilities. Note that the pathway for adding an existing observation (right-hand branch at the second node) will also be especially useful for non-TDA science applications, including archival work.}
    \label{fig:goats_workflow}
\end{figure}

The typical workflow that a user is expected to undertake with GOATS is shown in Fig.~\ref{fig:goats_workflow}. The following sections describe the software implementation for each step in the workflow. In addition to integrating the MMA/TDA services of NOIRLab/Gemini with the TOM-Toolkit base library, the development of GOATS includes several enhancements over the latter, pertaining to responsiveness, asynchronous operation, cybersecurity, and authentication as detailed below.

\subsection{Target Ingestion}\label{sec:antares}
GOATS provides multiple options to ingest targets (both time-domain and static sources) by interfacing with alert brokers and catalog services.

\begin{figure*}
    \centering
    \includegraphics[width=0.45\linewidth]{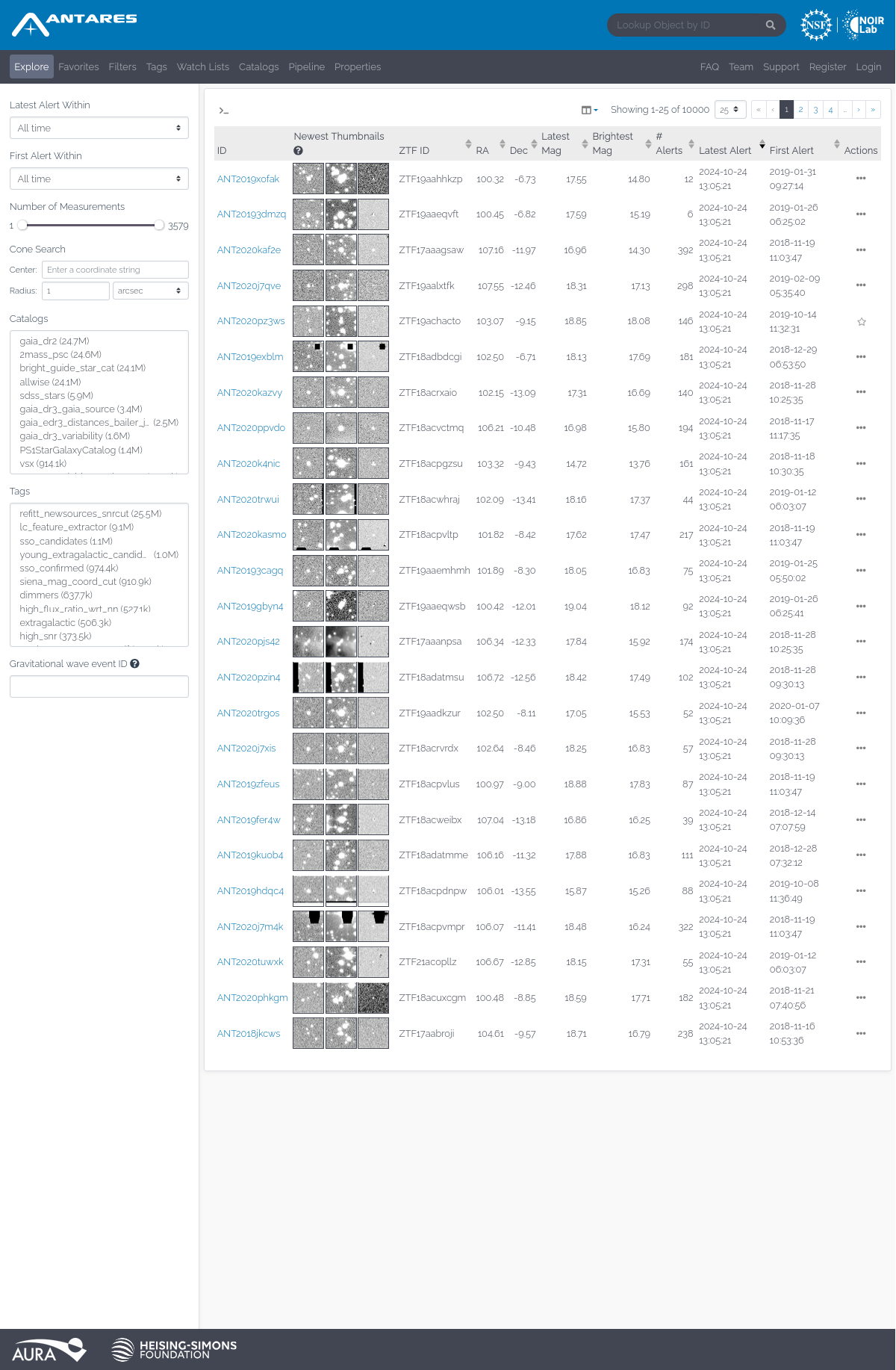}\hfil\includegraphics[width=0.45\linewidth]{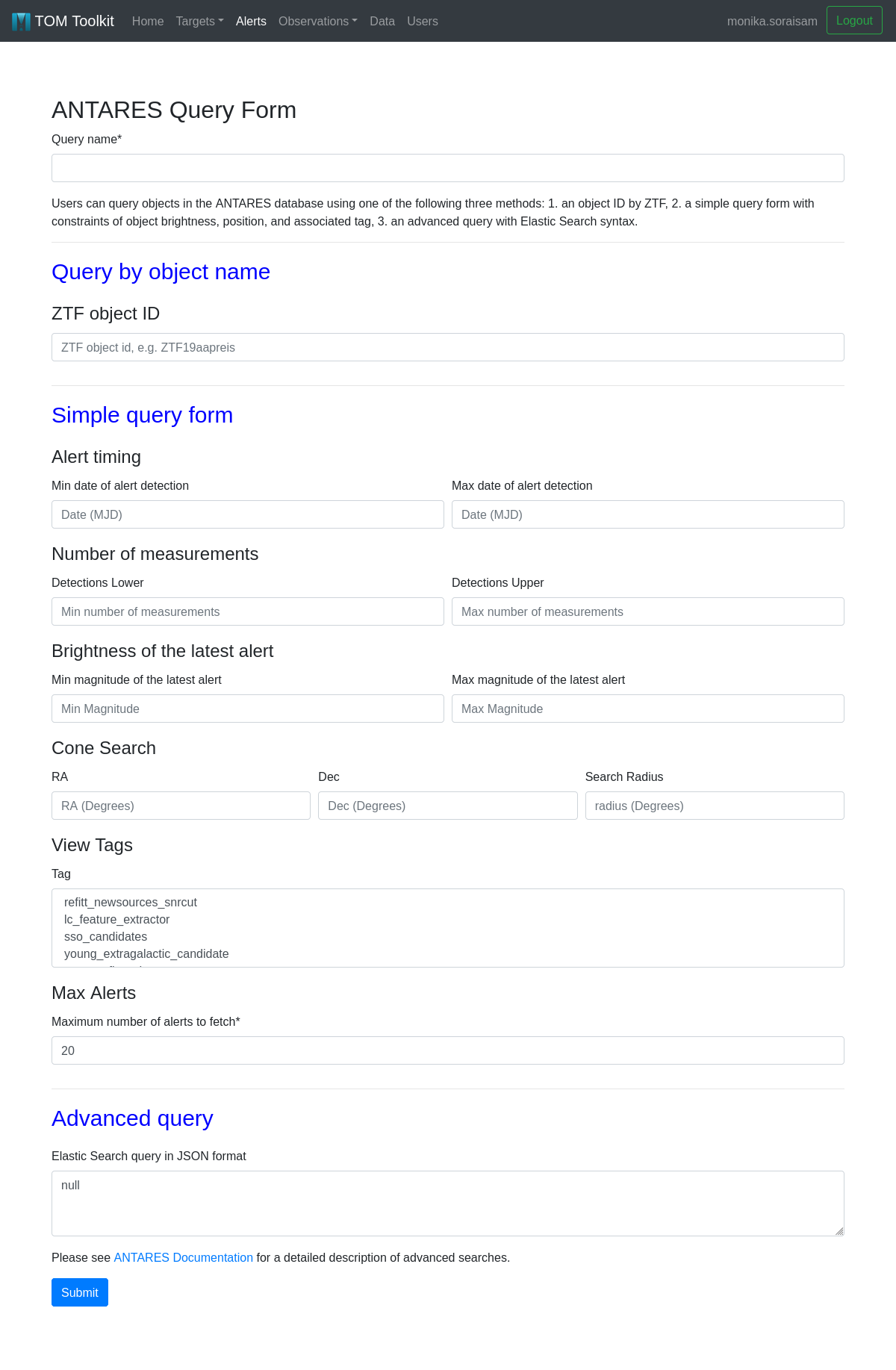}
    \caption{{\it Left}: ANTARES portal displaying controls for its available functionalities. {\it Right}: Query form rendered by the ANTARES TOM-Toolkit plugin, \texttt{tom-antares}.}
    \label{fig:tom_antares}
\end{figure*}

The ANTARES broker, developed by NOIRLab in collaboration with the University of Arizona, is one of the leading community brokers in operation and has also been selected to receive the full alert stream from Rubin/LSST. 
It offers multiple user interfaces, including Python client libraries to develop science filters to run in real time and a feature-rich web portal (cf.~Fig.~\ref{fig:tom_antares} left panel) that allows users to register watchlists of objects and receive Slack notifications of activity, and perform both simple and complex queries of alerts with the added values from ANTARES processing. 
An existing TOM-toolkit plugin for ANTARES (\texttt{tom-antares}) is available; however, it provides only limited support for querying the ANTARES alert database (see the right panel of Fig.~\ref{fig:tom_antares}).

\begin{figure*}
    \centering
    \includegraphics[width=0.6\linewidth]{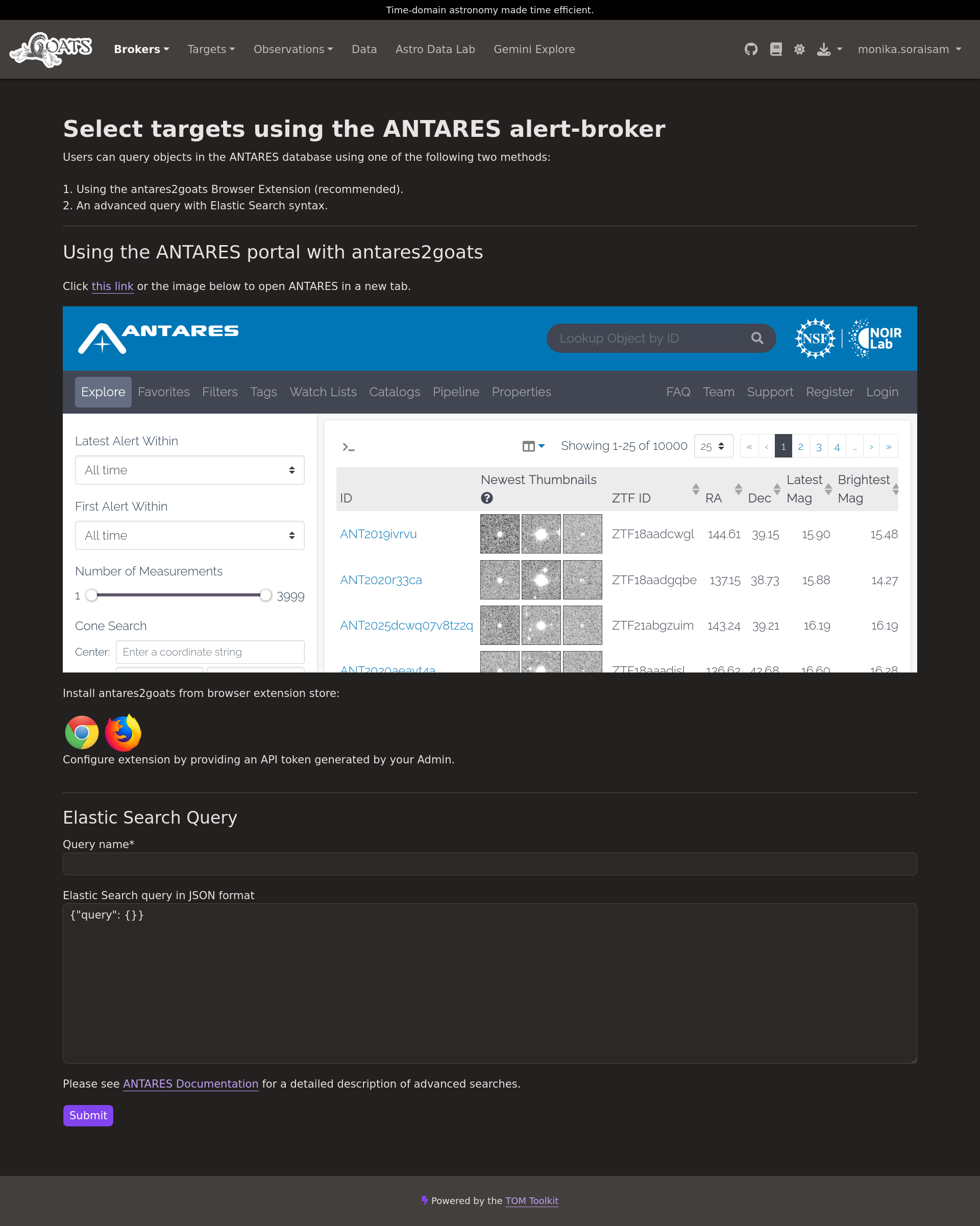}\\
    \vspace{2em}
    \includegraphics[width=0.6\linewidth]{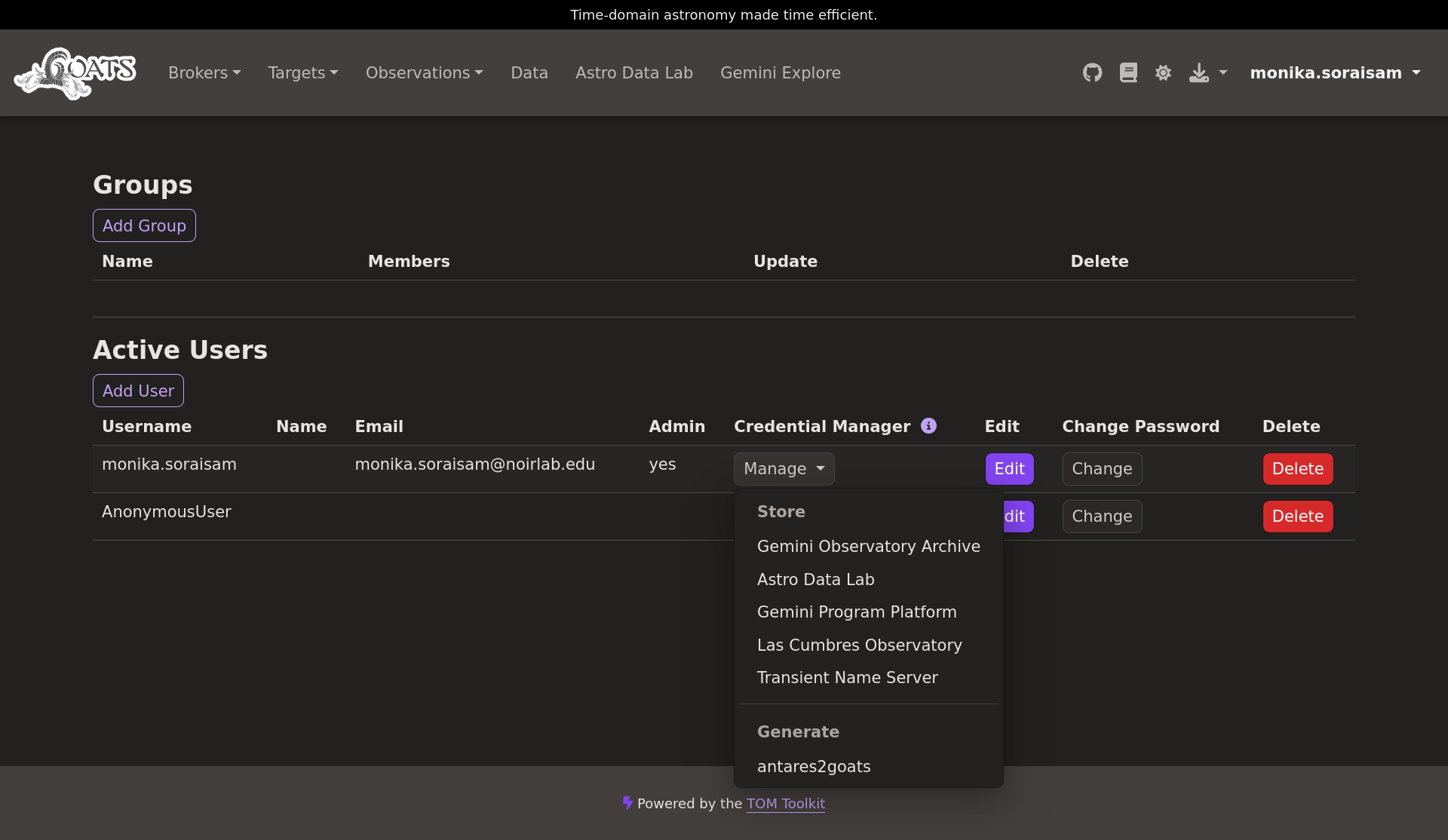}
    \caption{{\it Top}: Enhanced target selection capability implemented in GOATS using a browser extension to the ANTARES portal, with the additional option to use Elastic Search queries. {\it Bottom}: GOATS admin page for managing authentication to various services, including generating the token required to authenticate the \texttt{antares2goats} extension.}
    \label{fig:ant2goats}
\end{figure*}

\begin{figure*}
    \gridline{\fig{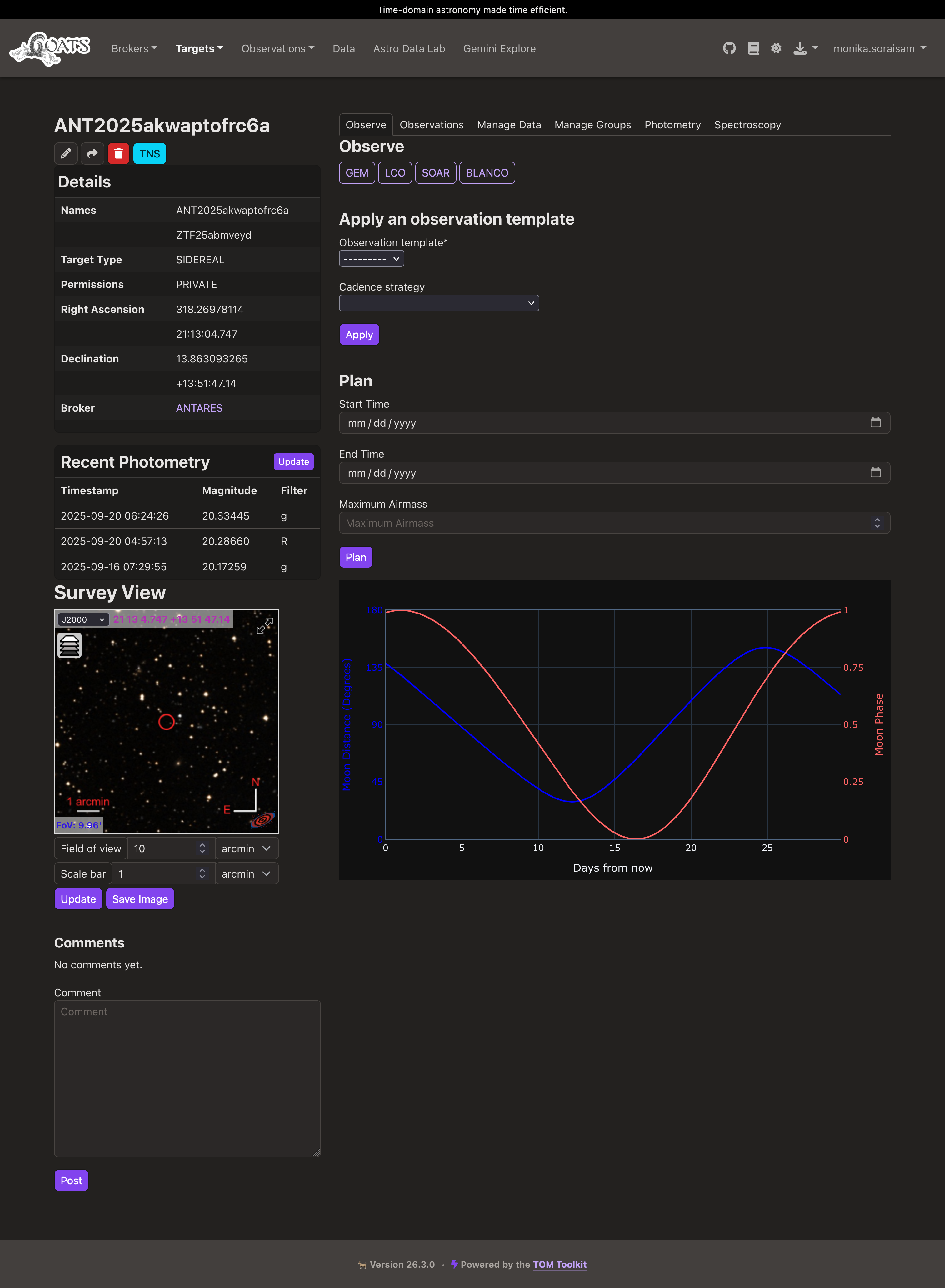}{0.55\linewidth}{}
            \raisebox{0.2\height}{\fig{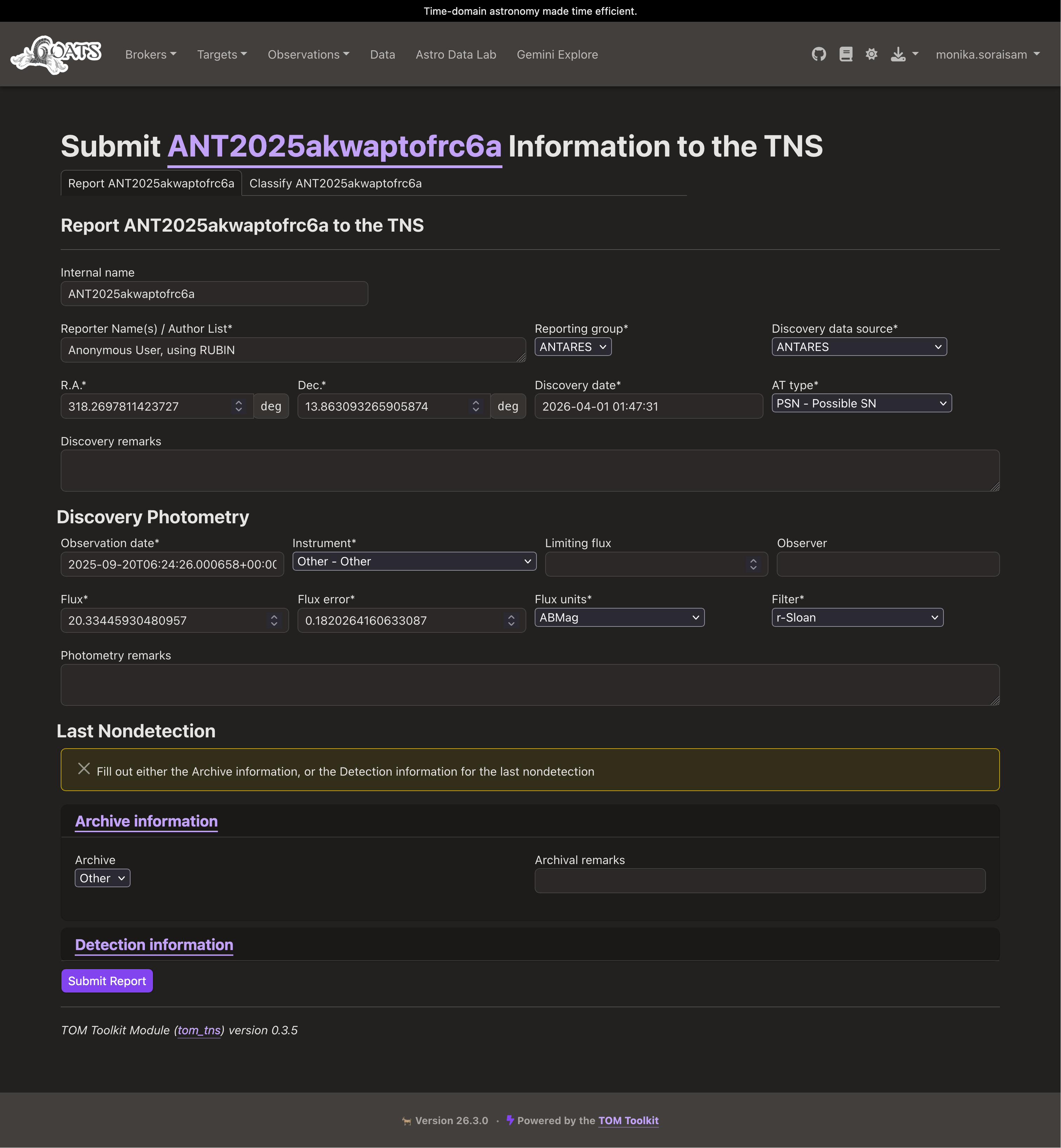}{0.45\linewidth}{}}
    }
    \caption{{\it Left}: Example target detail page on GOATS. The tabs to the right of the target name provide access to a variety of operations -- {\it Observe} allows users to trigger observations on AEON facilities; {\it Observations} displays and allows users to add existing observations of the target from supported facilities; {\it Manage Data} enables management of data collected for the target; {\it Manage Groups} lists and allows users to edit the group(s) to which the target belongs (if the user has grouped it with other targets); and {\it Photometry} and {\it Spectroscopy} can be used to display the corresponding data products. {\it Right}: Transient Name Server submission page on GOATS for reporting discovery of the target. Users can access this page by clicking the \texttt{{\bf TNS}} button in the left panel (highlighted in a teal box just below the target name).}
    \label{fig:target_page}
\end{figure*}

We have developed a browser extension, \texttt{antares2goats} (for Firefox{\footnote{\url{https://addons.mozilla.org/en-US/firefox/addon/antares2goats/}}} and Chrome{\footnote{\url{https://chromewebstore.google.com/detail/antares2goats/nmnbkpfjnpachfajklpjimbdpkoebcba}}}), through which GOATS integrates seamlessly with the ANTARES portal. This extension enables users to transfer search queries or target information (e.g., co-ordinates, target name) directly to their GOATS instance with the click of a button, without manually typing or copying information (Fig.~\ref{fig:ant2goats}), resulting in a dedicated target detail page (Fig.~\ref{fig:target_page}). The light curve of the target from ANTARES is also automatically ingested into GOATS (see Section~\ref{sec:analysis}).

We have incorporated token-based authentication, allowing users to generate tokens directly within GOATS, to ensure secure interactions with the extension. Users can thus leverage the full capabilities of the ANTARES portal. This significantly enhances workflow efficiency compared to traditional TOM-Toolkit broker plugins. Although not as powerful as using the ANTARES portal, GOATS also integrates plugins for other brokers such as ALeRCE, to enable users to harvest follow-up targets from multiple sources.

Users can also manually input target information into GOATS or upload a CSV file for multiple targets. Additionally, target information can be retrieved from catalogs such as SIMBAD, NED, Transient Name Server (TNS), and JPL Horizons. 

GOATS further allows users to submit discovery and/or classification information directly from the target detail page to TNS (see Fig.~\ref{fig:target_page}). 

\begin{figure*}
    \gridline{\fig{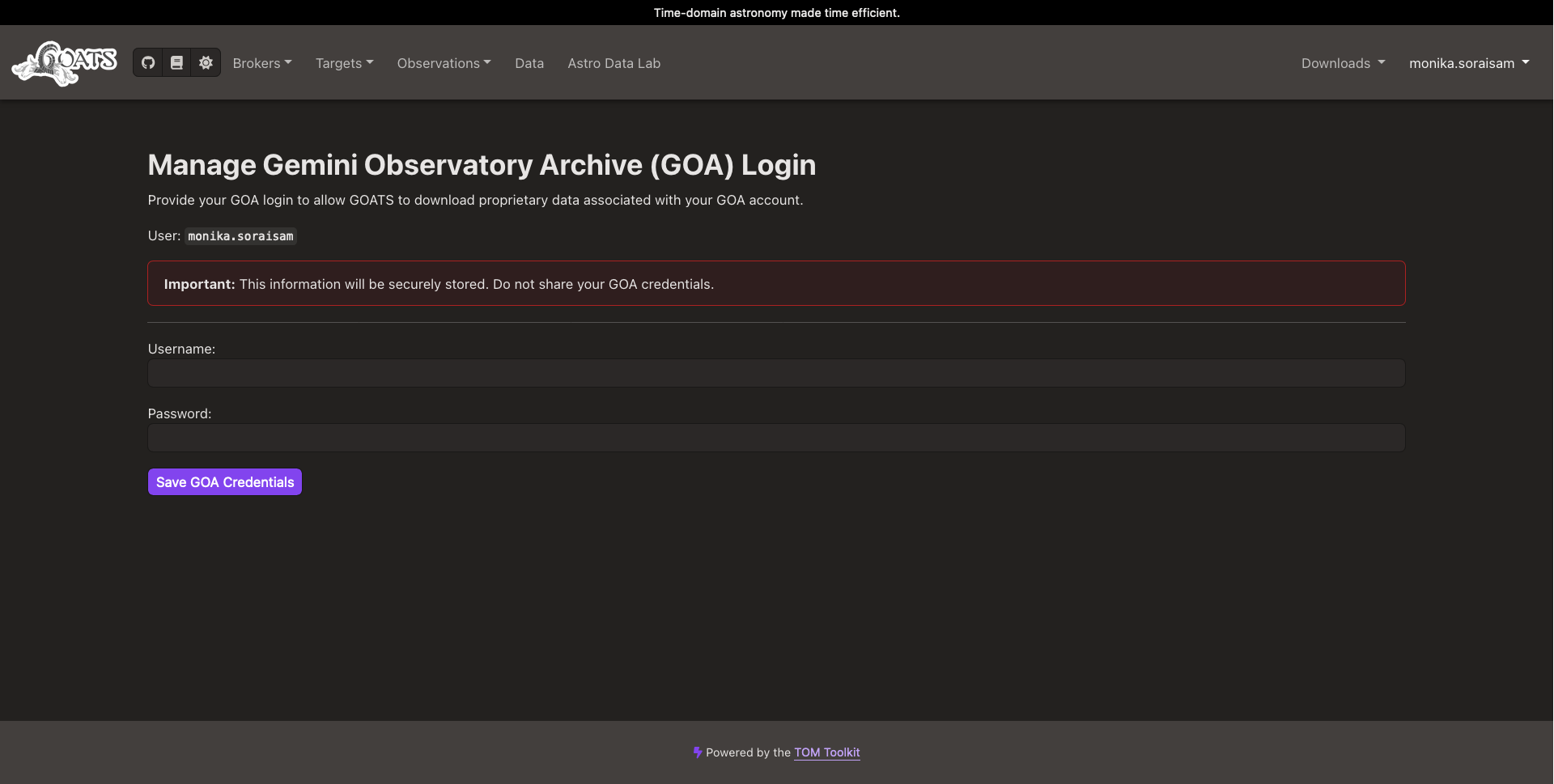}{0.6\linewidth}{(a)}}
    \gridline{\fig{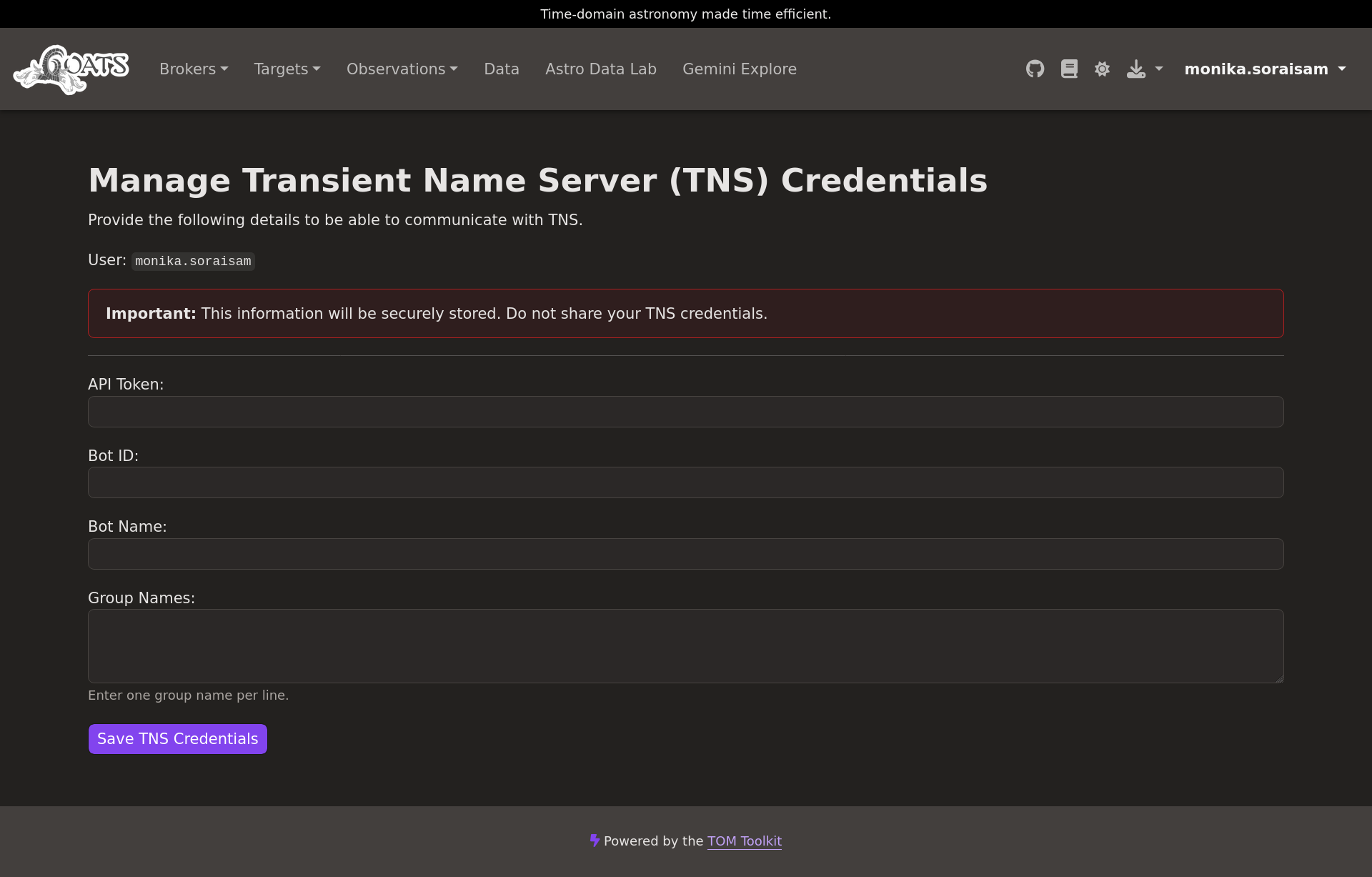}{0.6\linewidth}{(b)}}
    \gridline{\fig{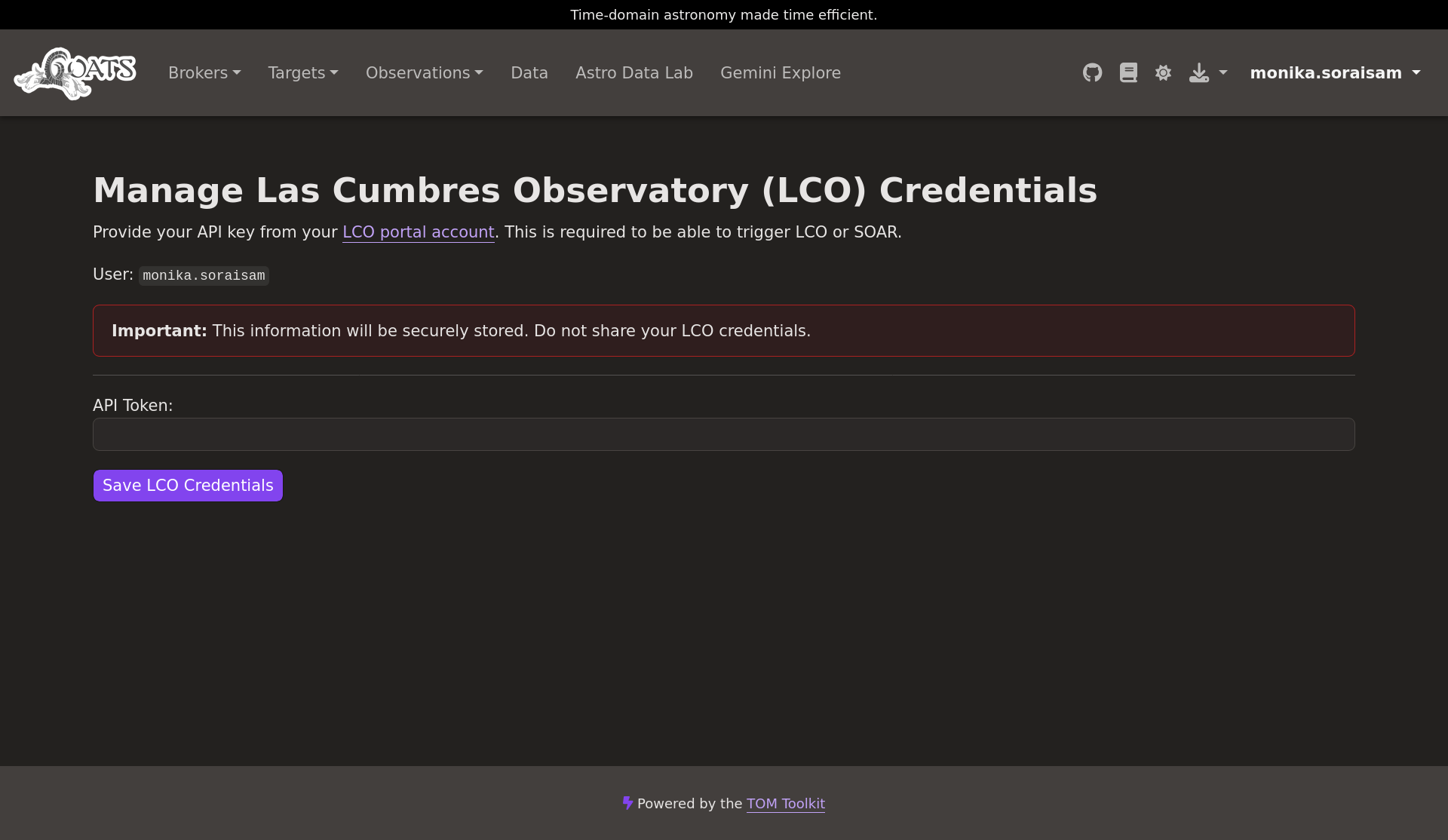}{0.6\linewidth}{(c)}}
    \caption{Credential management page for {\it (a)} Gemini Observatory Archive, {\it (b)} Transient Name Server (TNS), and {\it (c)} Las Cumbres API key. The latter is used by both SOAR and Blanco facilities. Users can navigate to these pages by clicking the corresponding service in the menu under \texttt{Credential Manager} of GOATS (see bottom panel of Fig.~\ref{fig:ant2goats} ). By registering credentials on these pages, users will be able to download their Gemini proprietary data, submit discovery and classification report of transients to TNS, and trigger Las Cumbres, SOAR and Blanco.}
    \label{fig:admin_pages}
\end{figure*}

\begin{figure*}
    \centering
    \includegraphics[width=0.8\linewidth]{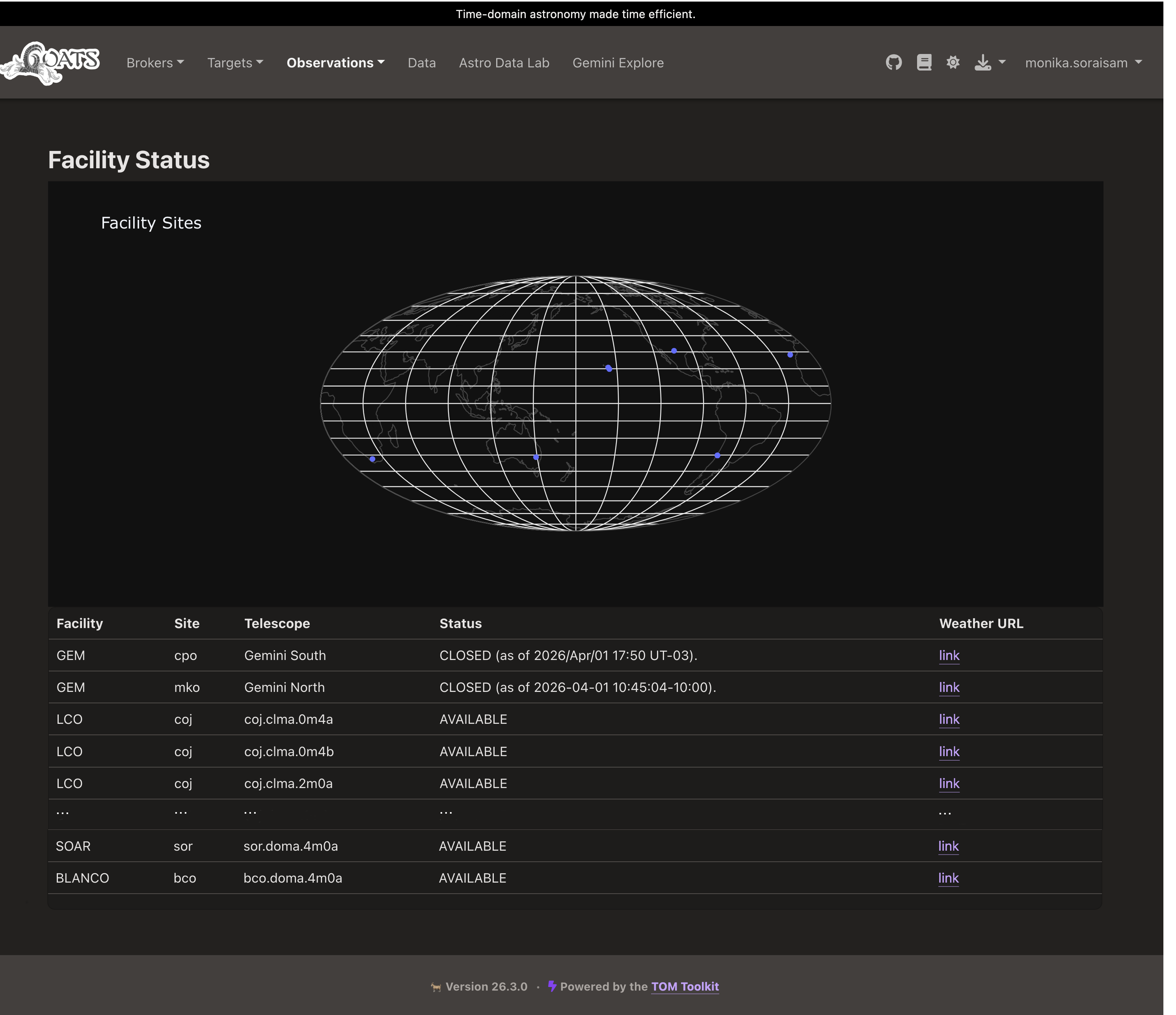}
    \caption{Facility status page in GOATS. Users can access this page under the \texttt{\bf Observations} tab in the top navigation bar to view the real-time operational status of both Gemini telescopes as well as other AEON facilities.}
    \label{fig:gem_status}
\end{figure*}

\subsection{Triggering Gemini: interfacing with the Gemini Program Platform}\label{sec:gpp}
There is active development to modernize the Observatory Control System (OCS) software at Gemini to replace its legacy system that has been in use for nearly 20 years but has become increasingly difficult to maintain and cannot support many required new features. The OCS supports both proposal submission and observation preparation by the user community, as well as planning and execution of observations by Gemini staff. An existing Gemini TOM-Toolkit plugin, built on the legacy OCS API, enables users to submit observation requests but it offers only limited functionality. Users must still prepare observations in advance with the OT during Phase II, and only a small subset of observation parameters can be configured through the API. This limitation is particularly problematic for MMA/TDA ToO programs. For example, when a transient is identified, users may wish to update the central wavelength of a spectroscopic observation or modify the sequence of exposures---capabilities not supported by the current Gemini API/TOM-Toolkit plugin, requiring them to fall back on the OT.

The Gemini Program Platform (GPP) is the key piece in the effort to modernize the OCS. Its goal is to improve observing efficiency, streamline science operations, ensure flexibility and responsiveness required in the era of MMA/TDA, and avoid code obsolescence (thus making the code maintainable and scalable). Among other capabilities, GPP includes an automated scheduler, a web interface called {\it Explore} for proposal submission and observation preparation oriented toward the science goal rather than low-level instrument configurations, a new observing database with improved query access, secure web APIs to all its services, etc. {\it Explore} is now online{\footnote{https://explore.gemini.edu/}} and expected to be in use by the community for preparing Gemini proposals in 2027. {\it Explore} is already linked in the GOATS interface (at the top navigation bar), from which users can launch it directly.

\begin{figure*}
    \centering
    \includegraphics[width=0.7\linewidth]{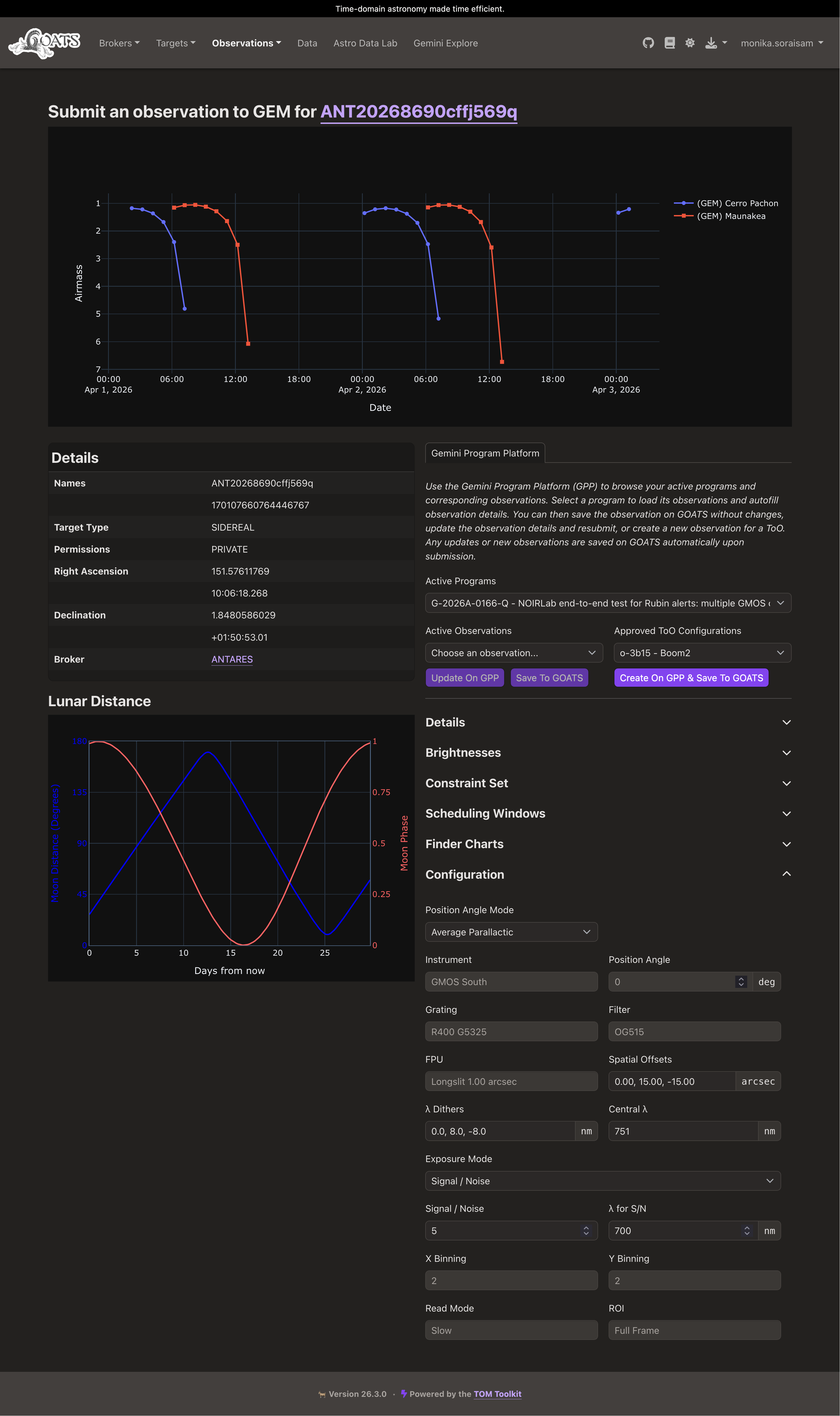}
    \caption{Page to submit an observation to Gemini using its new OCS, i.e., Gemini Program Platform. Users can navigate to this page by clicking \texttt{\bf GEM} under \texttt{\bf Observe} in the target detail page (see Fig.~\ref{fig:target_page}).}
    \label{fig:trigger_gemini}
\end{figure*}

At the time of writing, support for the most popular instrument and its mode (GMOS longslit spectroscopy) has been incorporated in GPP and is undergoing active testing. We have fully implemented the connection between GPP and GOATS such that Gemini can be triggered directly from GOATS. It is to be noted that with GPP/Explore, users will prepare observations{\footnote{For ToO observations, this entails setting up the instrument configurations to be used and the observing constraints during proposal submission itself instead of a Phase II process.}} and submit them together with the proposal for review. Once granted time by the Time Allocation Committee, users will be able to access their approved Gemini program(s) on GOATS and select a ToO configuration and edit as appropriate to create a new observation for their target as shown in Fig.~\ref{fig:trigger_gemini}. The form rendered in GOATS is a lite version of the observation creation page in Explore, but covers all the fields relevant for ToO triggering so users do not need to leave the GOATS interface. Once triggered, GPP will automatically generate the optimal observation sequence including calibrations and the newly created  observation is automatically reflected in the Explore website.

Additionally, users can add and work with existing Gemini observation IDs for their targets in GOATS (see Fig.~\ref{fig:observation} left panel; Section~\ref{sec:goa}). The real-time status of the Gemini telescopes along with the link to the weather on the summit are also accessible on GOATS (see Fig.~\ref{fig:gem_status}).

\subsubsection{Triggering other AEON facilities: SOAR, Blanco and Las Cumbres Observatory}\label{sec:aeon}
\begin{figure*}
    \centering
    \includegraphics[width=0.6\linewidth]{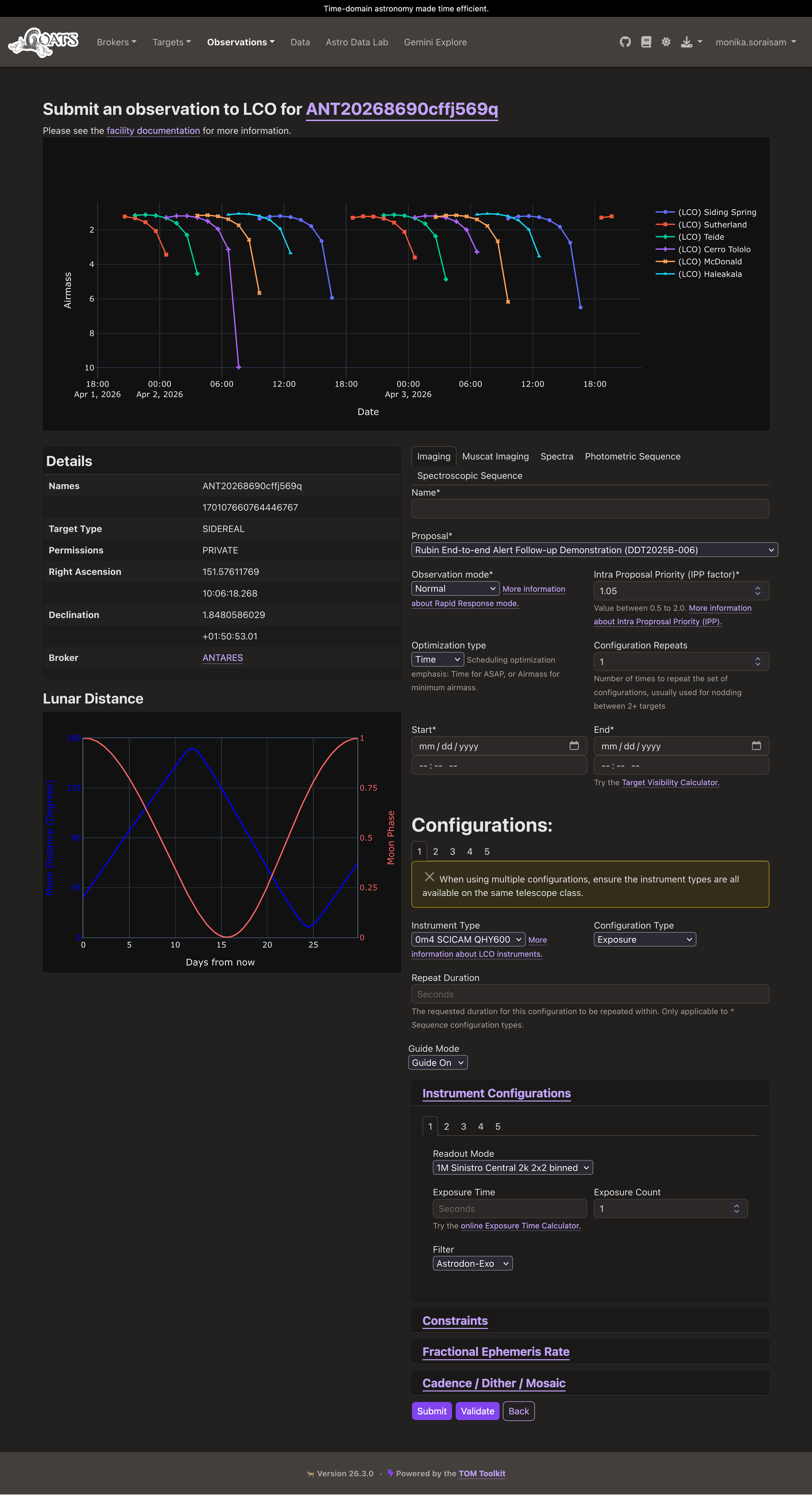}
    \caption{Page to submit an observation to Las Cumbres. The observation triggering page for SOAR and Blanco look similar. Users can navigate to this page by clicking \texttt{\bf LCO} under \texttt{\bf Observe} in the target detail page (see Fig.~\ref{fig:target_page}).}
    \label{fig:trigger_aeon}
\end{figure*}

The TOM-Toolkit base library underlying GOATS already includes support for triggering SOAR, Blanco (with the NEWFIRM instrument), and Las Cumbres, so GOATS can be used to trigger them as well. The former two facilities use the same scheduling software infrastructure as Las Cumbres. GOATS extends this capability with enhancements for ease of use. In particular, its Credential Manager (Fig.~\ref{fig:ant2goats} bottom panel) provides a user-friendly interface to input and store the required Las Cumbres API key, which users can obtain from their Las Cumbres portal{\footnote{\url{https://observe.lco.global/}}} account (see Figs.~\ref{fig:admin_pages} and \ref{fig:trigger_aeon}). The same key will also be used by GOATS for authenticating SOAR and Blanco triggers.     

\subsection{Interfacing with Gemini Observatory Archive}\label{sec:goa}
The Gemini Observatory produces a diverse data set from its large suite of optical/infrared instruments (including visitor instruments) hosted across two 8-m class telescopes in the northern and southern hemispheres. The Gemini Observatory Archive (GOA) provides user access to these data. Given the highly configurable nature of Gemini instruments, finding the appropriate calibration files for a given science dataset can be non-trivial. GOA solves this issue through its calibration module, which automatically locates the calibration data associated with each science dataset and presents both in the search results. GOA is deployed on Amazon Web Services, thus ensuring high availability to its users. Furthermore, data are ingested in less than a minute of an observation's completion, which benefits MMA/TDA science requiring rapid response. 

\begin{figure*}
    \centering
    \includegraphics[width=0.45\linewidth]{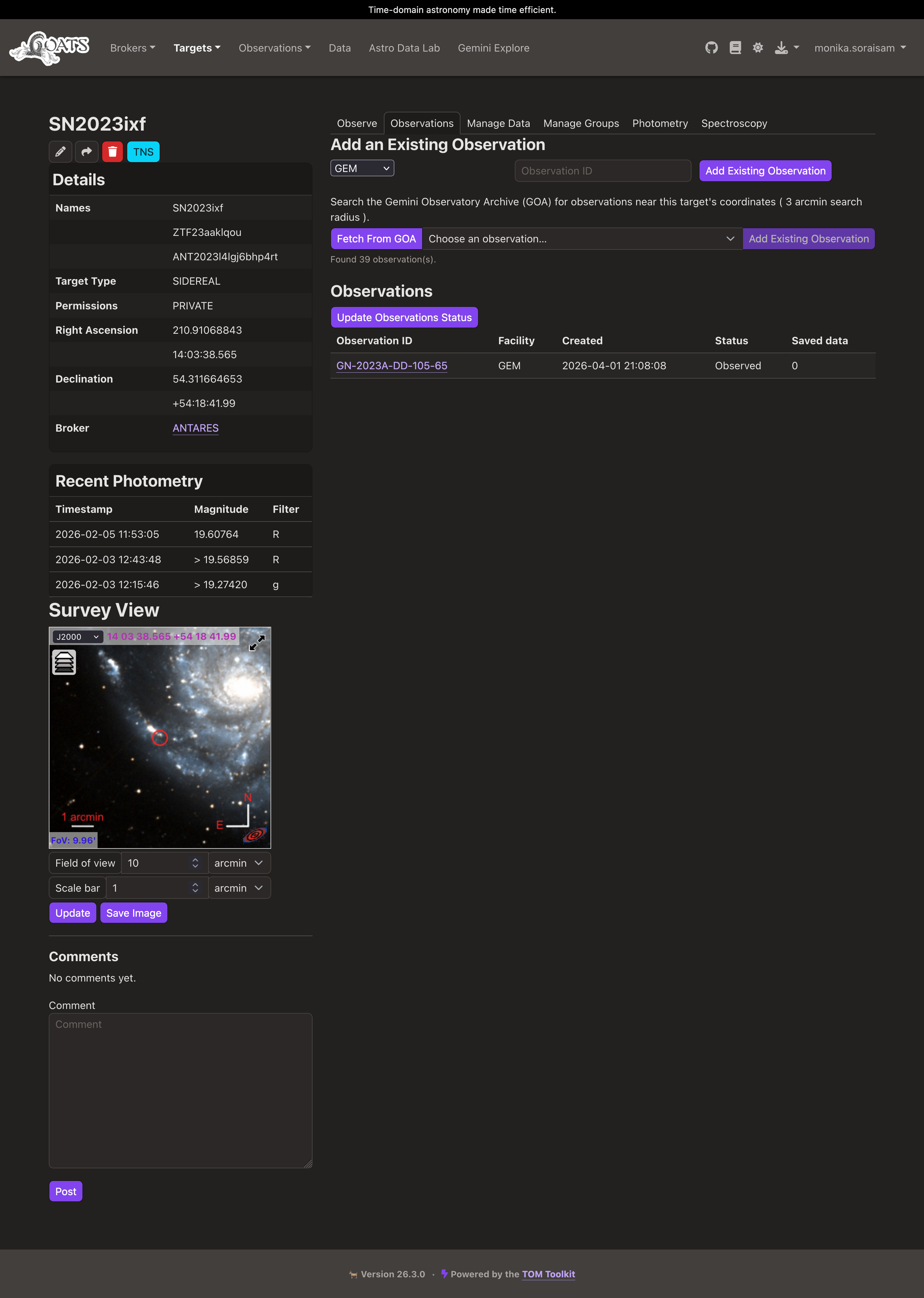}\hfil\includegraphics[width=0.435\linewidth]{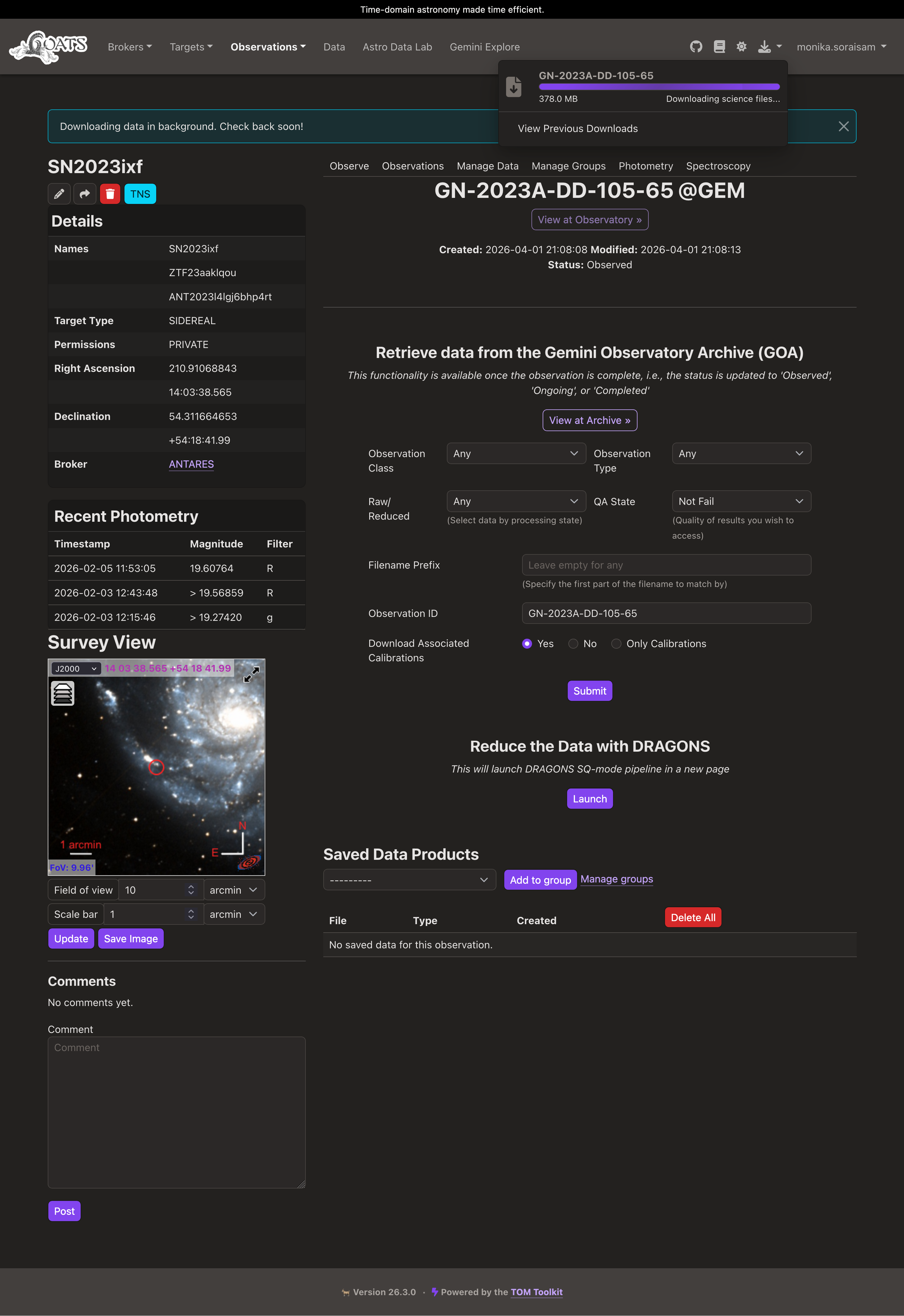}
    \caption{{\it Left}: {\it Observations} page for a given target in GOATS, which lists all the observations on GOATS associated with the target. Here users can also add existing Gemini observations of the target, either manually or automatically using the {\bf Fetch from GOA} button, and also obtain the status of the observation (shown here as {\it Observed}) by pressing the {\bf Update Observations Status} button. 
    {\it Right}: Observation detail page for the given ID. Through this page, users can download the data from GOA by simply hitting the {\bf Submit} button; the download progress can be monitored via the {\bf Downloads} button. By default, this will download both the science and associated calibration files, automatically unzip the files and list them under {\bf Saved Data Products}.}
    \label{fig:observation}
\end{figure*}

GOA also provides a RESTful API for programmatic access. GOATS uses this API to automatically fetch existing Gemini observation IDs for a given target and to download data directly from GOA without leaving the GOATS portal (Fig.~\ref{fig:observation}). By default, both science and associated calibration data for an observation are downloaded, though users can customize the download options within the portal. GOATS supports downloading both public and proprietary datasets; for the latter, users need only register their GOA login{\footnote{Currently GOATS does not support ORCID authentication for GOA.}} credentials once via the {\bf GOA Login} option in the admin page (see bottom panel of Fig.~\ref{fig:ant2goats} and Fig.~\ref{fig:admin_pages}). 

Data downloads are handled asynchronously using modern technologies such as Redis and Dramatiq. This allows users to download multiple Gemini observation datasets simultaneously without interrupting other operations on GOATS.

\subsection{Interfacing with DRAGONS -- the Gemini data reduction software}\label{sec:dragons}
Gemini provides DRAGONS \citep{DRAGONS}---a versatile, Python-based, open-source data reduction framework---for reducing data generated by the Observatory's myriad instruments. Development of DRAGONS prioritizes code reuse, thereby minimizing long-term maintenance cost. Among its various features, DRAGONS performs automated association of processed calibration files during the reduction and provides browser-based interactive tools, driven by \texttt{bokeh} at the backend, for users to optimize their data reduction.

A unique feature of GOATS is that it fully integrates DRAGONS into its software stack, thus enabling users to reduce Gemini data on the GOATS interface. In fact, this integration transforms DRAGONS into a web application. GOATS also includes several additional features (see below) making data reduction with DRAGONS on its interface more user-friendly than using the native DRAGONS API or command line.

\begin{figure*}
    \gridline{
        \raisebox{0.8\height}{\fig{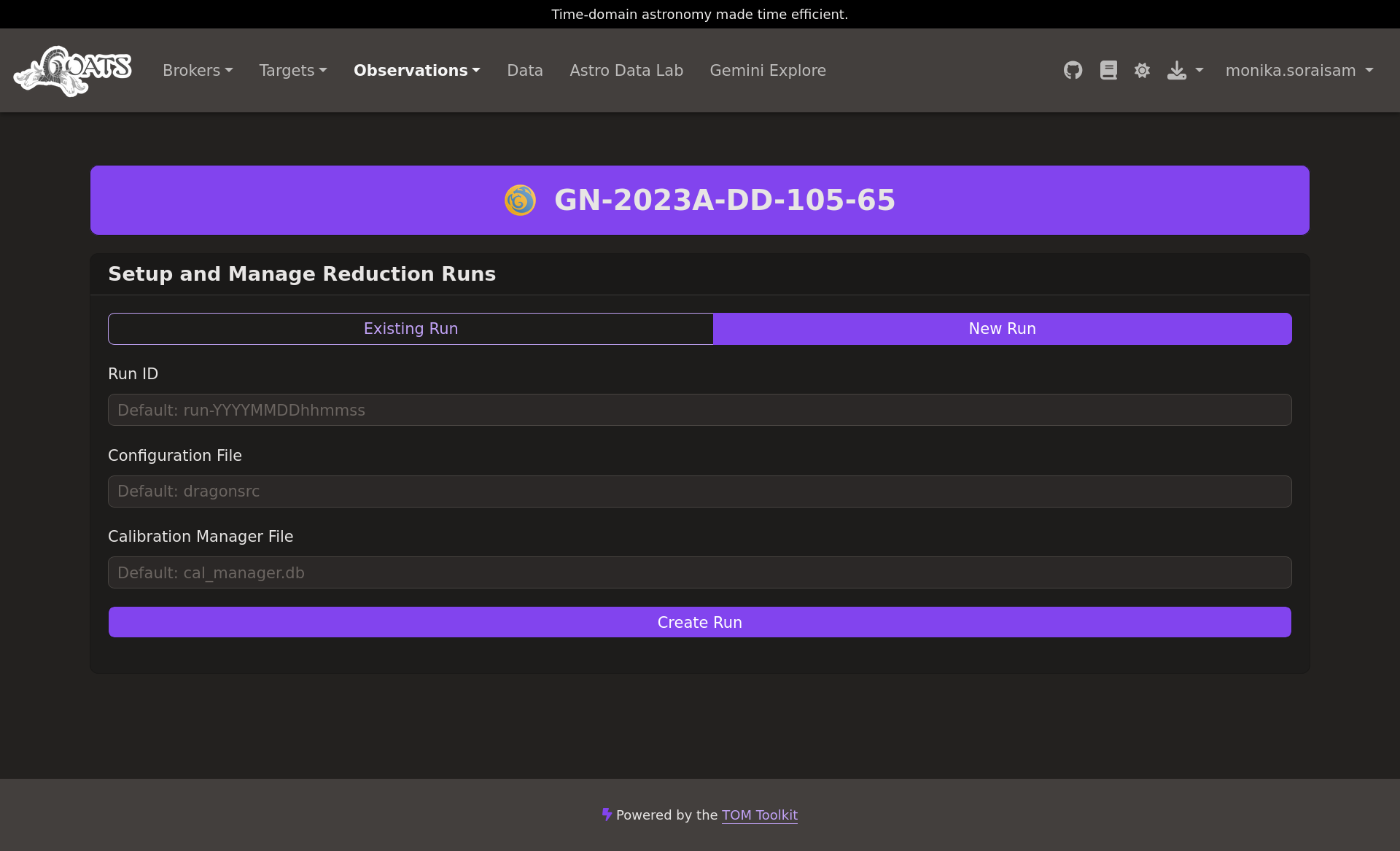}{0.45\linewidth}{}}
        \fig{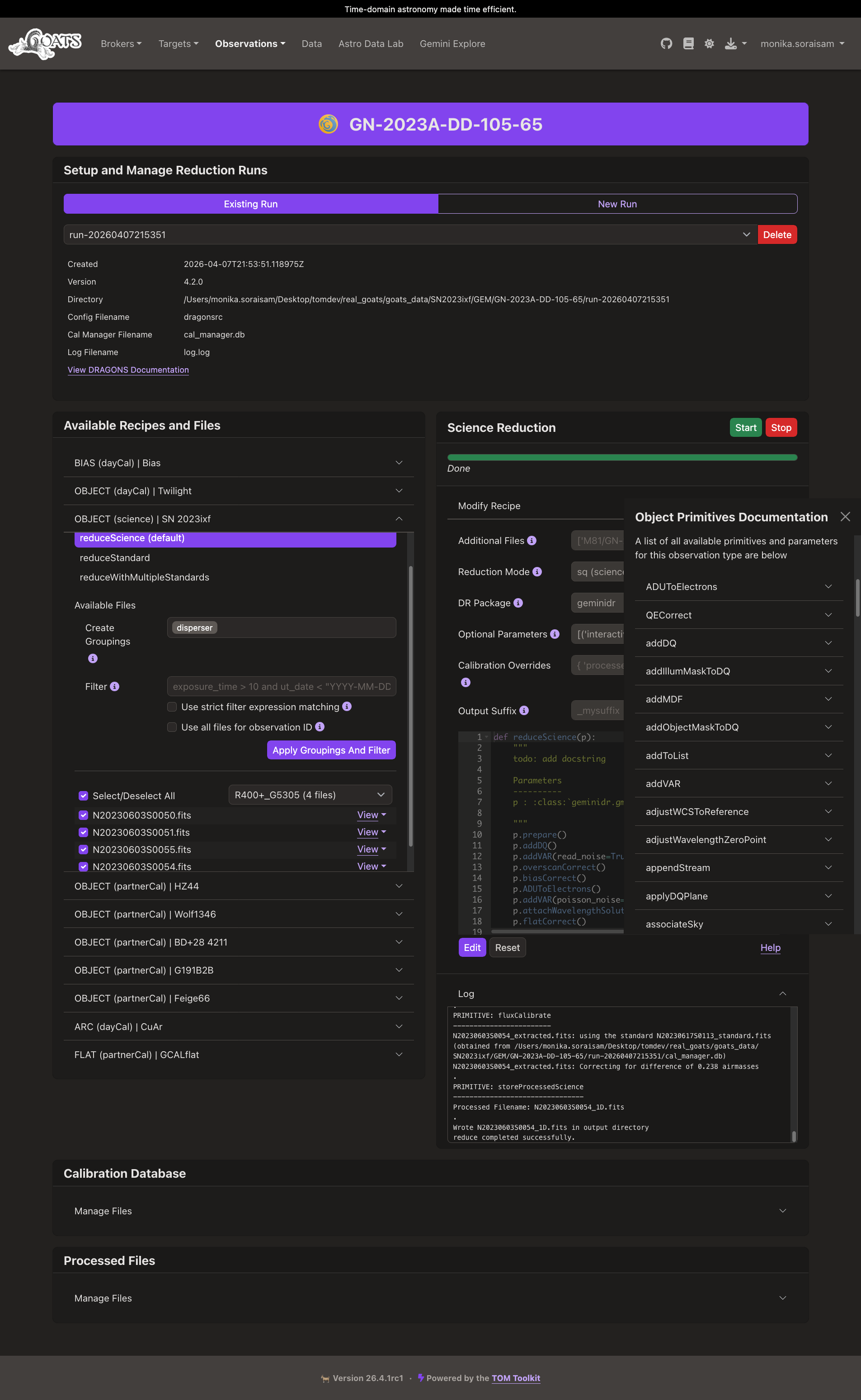}{0.55\linewidth}{}
    }
    \caption{{\it Left}: Setup page for configuring a new DRAGONS reduction session in GOATS for a given Gemini observation ID. 
    {\it Right}: DRAGONS reduction page for a selected reduction session in the GOATS portal, highlighting features designed to make the reduction more user-friendly. These include automated sorting of files into observation types, access to all available reduction recipes, and tools for further grouping and filtering files. The right-hand side of the page displays the source code for the chosen reduction recipe, with options to edit or customize it directly within GOATS. Documentation for the functions used in the reduction recipe is also generated dynamically and is accessible directly from the interface.}
    \label{fig:dragons}
\end{figure*}

Users can {\it launch} the DRAGONS application directly from the detail page of a given observation in GOATS (cf.~Fig.~\ref{fig:observation} right). This opens the application in a separate tab (see left panel of Fig.~\ref{fig:dragons}), where users are presented with the option to either setup a new reduction session -- for example, by editing the default session name, DRAGONS configuration file name, or calibration manager file name -- or to select an existing session for the observation. 

Once a session is chosen, the DRAGONS reduction page is displayed. Here, all files for the given observation ID are automatically sorted by observation type, with the available reduction recipes for each type listed alongside (Fig.~\ref{fig:dragons} right panel). Users can further group and/or filter files within an observation type, and the basic information of the run is accessible at the top of the page. 

After selecting a reduction recipe, users may customize the reduction code directly within the GOATS interface (Fig.~\ref{fig:dragons} right panel). Documentation for each function in the reduction recipe is available on the fly, integrated into the portal. Users can also manage (view, remove, etc.) output and processed calibrations files via the {\bf Processed Files} and {\bf Calibration Database} menus (visible near the bottom of Fig.~\ref{fig:dragons} right panel). Finally, the state of each session is preserved, allowing users to revisit them at any time.

\subsection{Interface with the Astro Data Lab science platform}\label{sec:datalab}
The Astro Data Lab \citep[henceforth DL;][]{DataLab2014,DataLab2020} is a science platform developed and managed by NOIRLab. It offers open access to public photometric and spectroscopic survey datasets and facilitates interactive and programmatic data discovery and retrieval. Users can perform SQL/ADQL database queries via IVOA's Table Access Protocol (TAP). Generous allocations are provided for remote file storage (VOSpace) and user-owned database tables (MyDB), which are located alongside DL's extensive data holdings. Additionally, the platform offers various services including cross-matching, image discovery, and image cutouts through the Simple Image Access protocol, as well as file services for survey data. Furthermore, a Jupyter notebook interface is available for data analysis in close proximity to the datasets. Sharing data tables, files, and workflows among DL users is also a core functionality.

\begin{figure*}
    \gridline{
        \fig{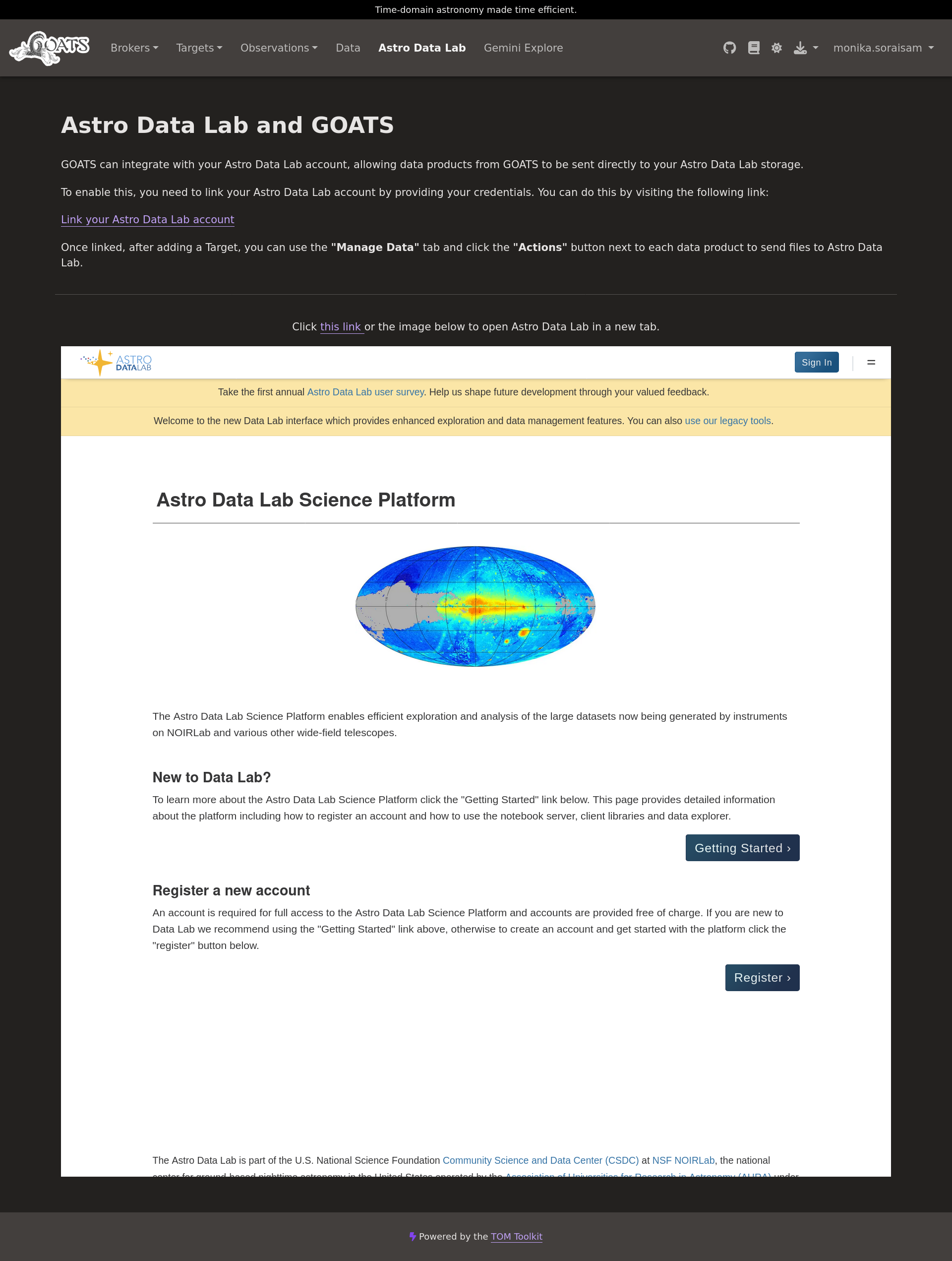}{0.5\linewidth}{}
        \fig{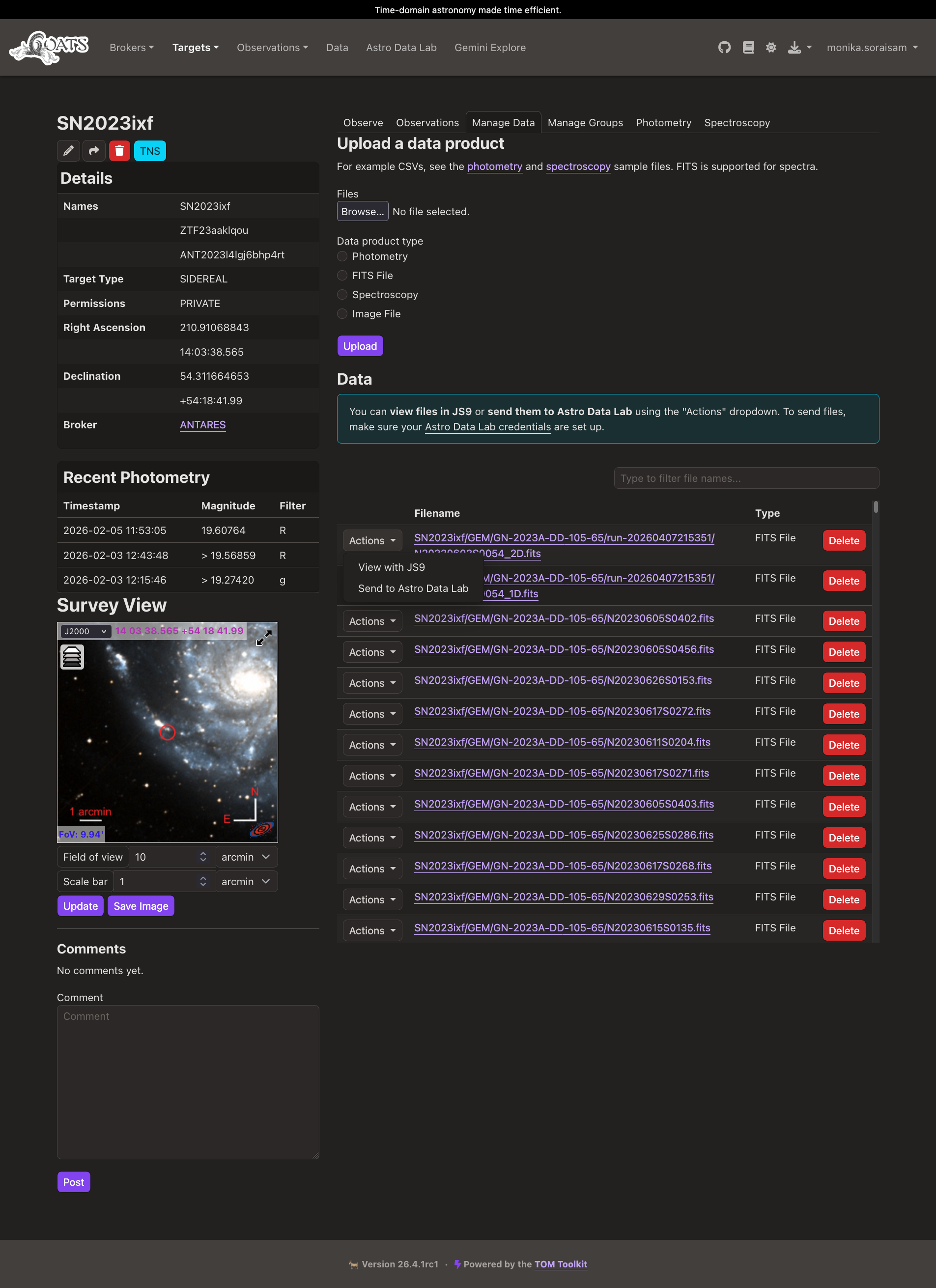}{0.485\linewidth}{}
    }
    \caption{{\it Left}: Astro Data Lab (DL) page on GOATS. By simply clicking, users can launch DL directly from GOATS. {\it Right}: The {\it Manage Data} page for a given target on GOATS provides an option (under the Actions button for a given file) to transfer data to the user’s DL account. This page aggregates all data available for the target.}
    \label{fig:datalab}
\end{figure*}

DL is directly connected with GOATS, allowing users to access their DL accounts through the GOATS interface (cf.~Fig.~\ref{fig:datalab}) and make use of the various data services offered by DL as noted above. Users can manage their DL authentication on the GOATS admin page (see Fig.~\ref{fig:ant2goats} bottom panel), in a manner similar to the GOA login process (Section~\ref{sec:goa}). With a single click (Fig.~\ref{fig:datalab} right panel), users can transfer data from GOATS to DL, perform data analysis using a Jupyter notebook on DL (including joint analysis with other DL-hosted data products), share data with collaborators, and even use the platform for long-term hosting of their data products.

\begin{figure*}
    \gridline{
        \fig{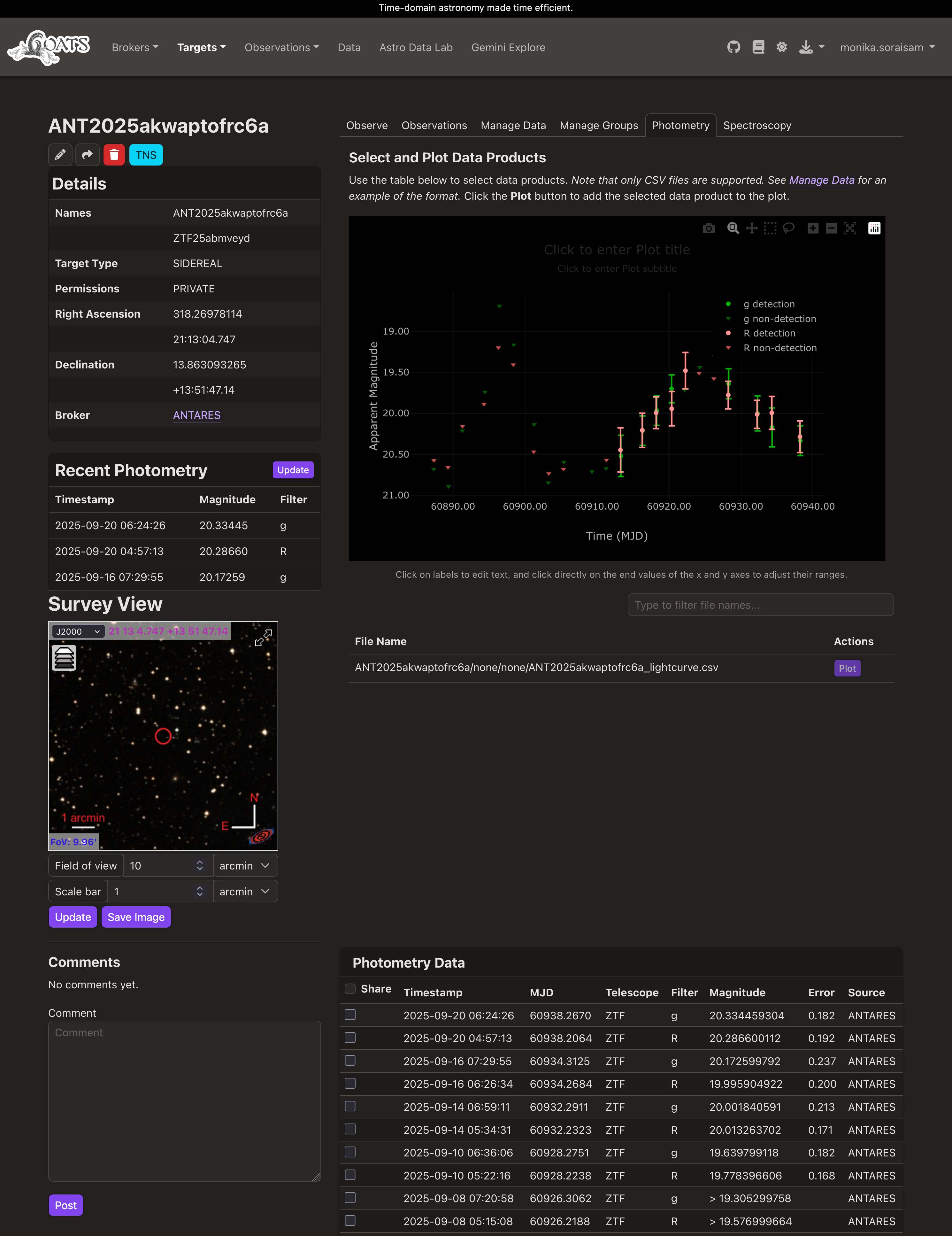}{0.5\linewidth}{}
        \fig{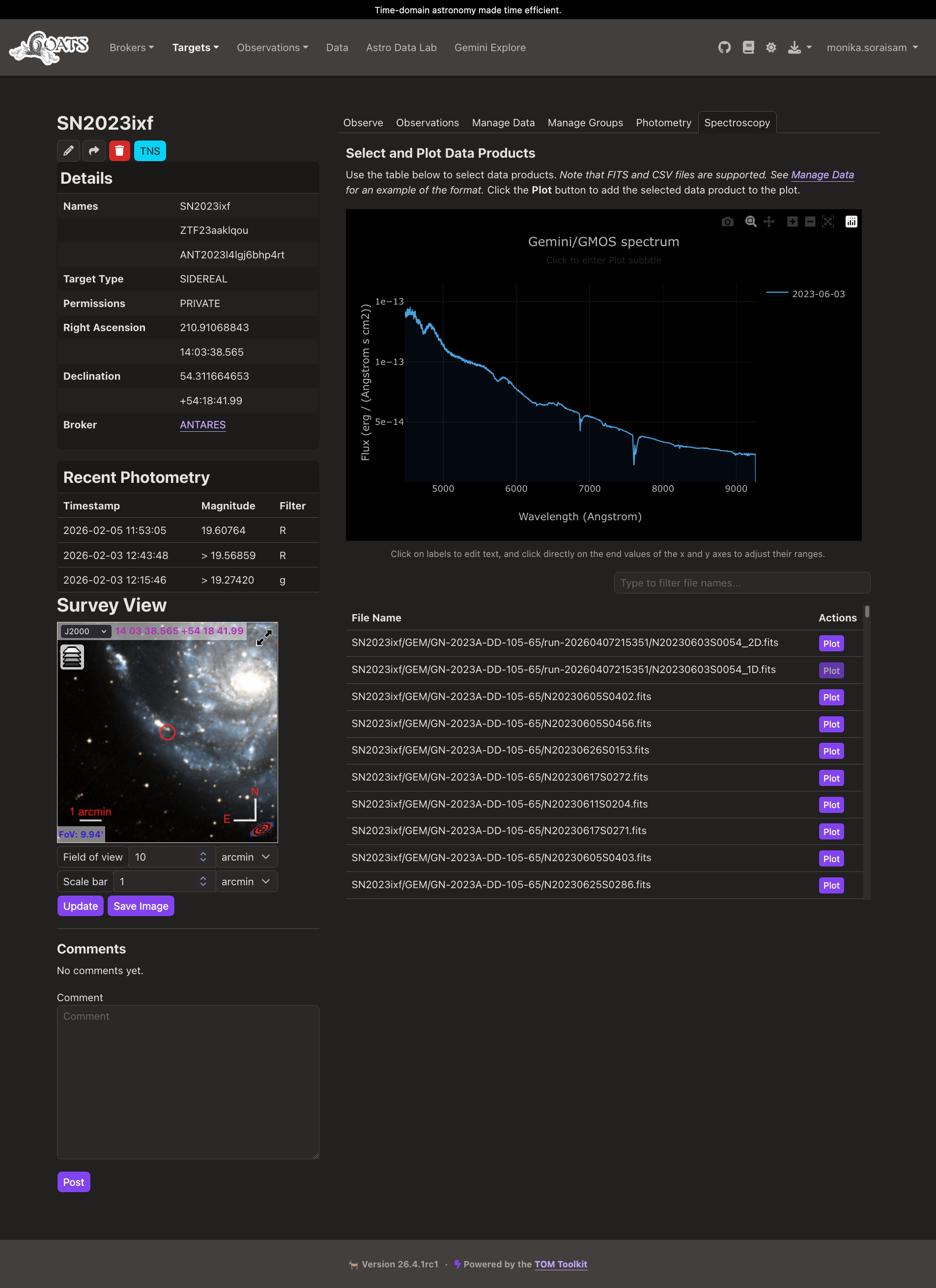}{0.474\linewidth}{}
    }
    
    \caption{{\it Left}: Photometry visualization page for a given target in GOATS, displaying its light curve ingested from ANTARES. {\it Right}: Spectroscopy visualization page in GOATS, showing the DRAGONS-reduced spectrum from GMOS R400 grating data for SN~2023ixf, obtained as part of observation ID GN-2023A-DD-105-65 (PI: Lotz).}
    \label{fig:vis}
\end{figure*}

\subsection{Data visualization and analysis 
}\label{sec:analysis}

GOATS supports visualization of data, both photometry and spectra for a given target (see Fig.~\ref{fig:vis}). This capability is provided by the TOM-Toolkit base library on which GOATS is built. As part of our ongoing development, we plan to extend this functionality to support data analysis, particularly for spectroscopic data. Users will be able to interactively identify spectral lines, determine redshifts, fit line profiles, etc.\ directly within the interface. These tools will be applicable to any spectrum displayed on the page, not limited only to Gemini data.

\subsection{Software availability}\label{sec:soft}
Accessibility and ease of use by the community are key development principles for GOATS. As a fully assembled TOM, users no longer need to configure any setting scripts or be familiar with web frameworks (e.g., Django). GOATS is distributed as a \texttt{conda} package; currently beta versions are available to the community. Installing and spinning up a GOATS instance requires executing just the following four simple commands:

\begin{verbatim}
    $ conda create -n your-env-name 
        python-version goats 
    $ conda activate your-env-name
    $ goats install 
    $ goats run 
\end{verbatim}

In the above, \texttt{your-env-name} is the name the user gives to their \texttt{conda} environment, and \texttt{python-version} is the version of Python to use in the environment{\footnote{For example, \texttt{conda create -n goats-env python=3.12 goats}
}}.

Detailed documentation is available at \url{https://goats.readthedocs.io/en/latest/}. Our code is open source and can be accessed at \url{https://github.com/gemini-hlsw/goats}. We have also made several contributions to the core TOM-Toolkit project as summarized in Appendix~\ref{sec:appendix}.

\section{Science verification: live end-to-end follow-up of Rubin alerts}\label{sec:science}
GOATS was recently used to successfully demonstrate NOIRLab's end-to-end capability for follow-up of Rubin/LSST alerts \citep{e2e}. 
The demonstration was carried out in collaboration with Las Cumbres on two consecutive nights, UT dates 2026-02-19 and 2026-02-20, with different alerts selected and followed up on different targets. The participating facilities of NOIRLab were Gemini (both telescopes), SOAR, and Blanco, and for Las Cumbres, its 1-m and 2-m telescope network.  
The back-to-back successful execution highlighted the robustness and reliability of the system. Using the follow-up observations, we classified the selected targets as supernovae of different types. The workflow during the test is summarized below.

We ran two science filters{\footnote{The filter codes can be accessed at \url{https://nsf-noirlab.gitlab.io/csdc/antares/devkit/reference/filters/}}} on ANTARES to flag alerts in real time that satisfied the following criteria:
\begin{itemize}
    \item Filter {\tt young\_rubin\_transients\_soraisam}: alert is associated with a ``transient" (see below), which is active within the past 30 days
    \item Filter {\tt rubin\_NovTest\_soraisam}: alert results in a blueward color evolution for the transient
\end{itemize}

The first filter was designed to cast a wider net. For both filters, several cuts on the alert properties were adopted to increase purity. This included the real-bogus score ($\rm{diaSource\_reliability} > 0.9$), extended-ness of the alert (to restrict to point-like objects), and light-curve amplitudes ($>0.5$~mag) specifically for those with all the detections confined to a 2~hr window, which helps get rid of new moving objects (without a match in the Solar System Object catalog) to some extent. In the absence of nearest-neighbor information in the alert packet (expected only after Data Release 1), we used the flux of the alert measured in the template image to determine whether the object is a ``transient", i.e., the template flux is consistent with noise. 

On the first night of our campaign, Rubin/LSST generated $\sim1$~million alerts and on the second night, $\sim2$~million alerts from the Deep Drilling Fields. The filtered alerts were tagged by ANTARES and streamed to a Slack workspace, which we monitored and vetted in real time to identify suitable follow-up candidates (considering criteria such as brightness and visibility from the available follow-up sites). The selected targets were automatically ingested into GOATS using our browser extension {\tt antares2goats}. Despite speed not being the primary goal of the demonstration, follow-up observations typically started within tens of minutes of the Rubin alert.

\begin{figure*}
    \centering
    \includegraphics[width=0.8\linewidth]{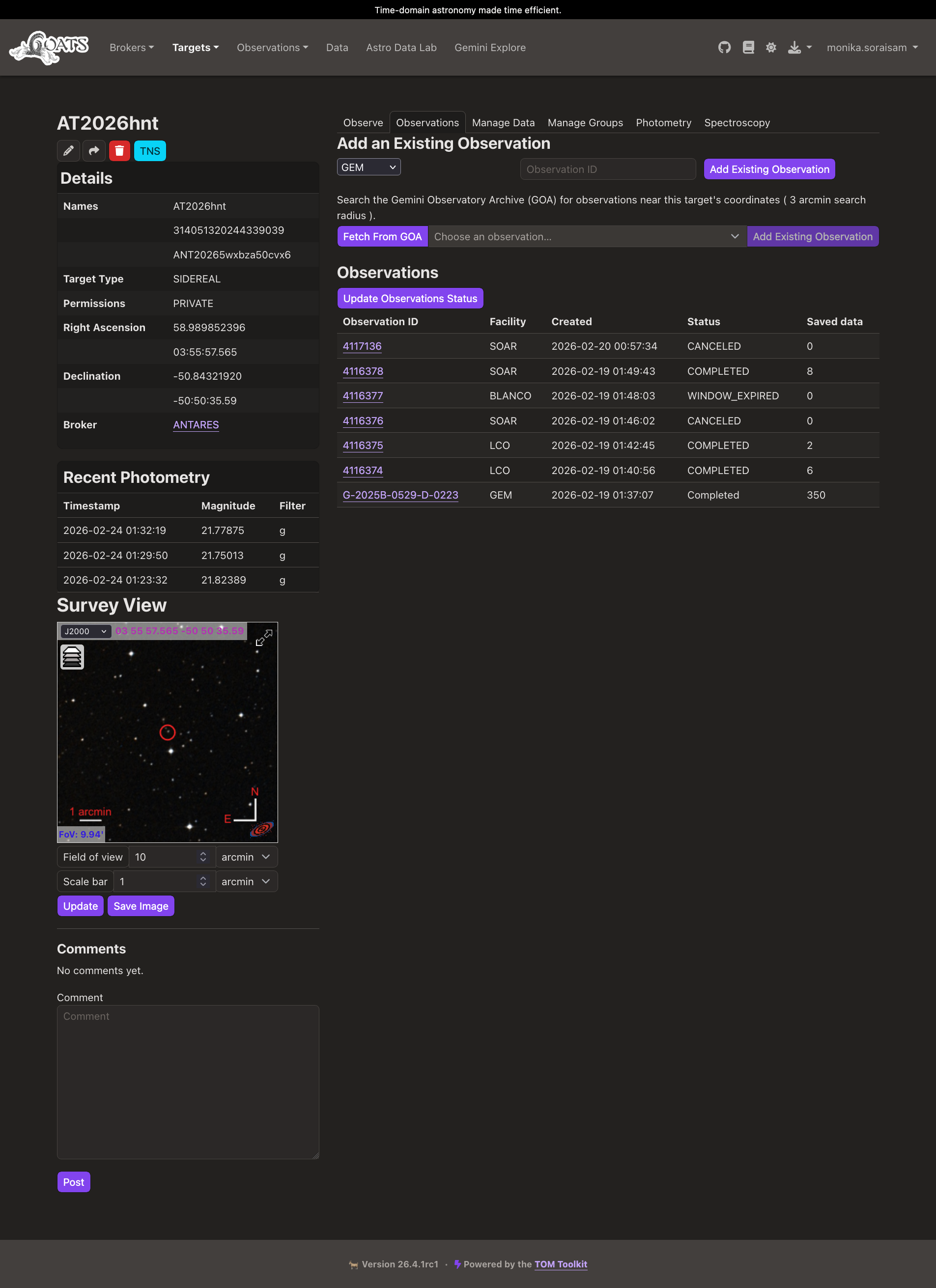}
    \caption{Summary of observations obtained for AT~2026hnt, which was the first target selected and triggered during the NOIRLab end-to-end demonstration. The list shows the timestamp (under the column {\bf Created}) when the observations were successfully triggered at various facilities and also the observation status (under the {\bf Status} column). The number of fits files retrieved and saved for each observation are listed under {\bf Saved data}. For Gemini, all raw files, including calibration files, were downloaded in GOATS for reduction with the integrated DRAGONS pipeline. For the other facilities, we only downloaded the reduced data products (automatically reduced by their respective pipelines). GOATS does not currently support data retrieval from Blanco. Note that the ANTARES id for this target shown in the image (i.e., ANT20265wxbza50cvx6) comes from its development database. Since the end-to-end runs were conducted before the public launch of Rubin/LSST alerts, ANTARES ingested the alerts using this development database.}
    \label{fig:2026hnt}
\end{figure*}

For example, we selected the first alert for follow-up (Rubin alert Id LSST-AP-DS-170028486043369574 from diaObject LSST-AP-DO-314051320244339039) within 7~min from the timestamp recorded in the alert packet (UTC 2026-02-19T01:16). This time interval encompasses ANTARES processing, the arrival of the Slack notification, and the vetting process, including the assessment of other alerts during that period. We then triggered rapid-response follow-up observations via GOATS on all participating facilities for this alert. A summary of those observations is shown in Fig.~\ref{fig:2026hnt} (see also Table~\ref{tab:e2e}). All the follow-up observations on the night of 2026-02-19 UT for this alert were successfully triggered within 34~min (first with Gemini within 21~min) of it being generated by Rubin. The triggering process only takes a few seconds, but we delayed it as we were vetting other candidates in the meantime. Without the delay, the first Gemini observation would have gone through within 8~min.

Note that a second alert (alert Id LSST-AP-DS-170032879130640394) from the same target was also flagged on the second night (2026-02-20 UT) and we had triggered SOAR (shown as observation Id 4117136 in Fig.~\ref{fig:2026hnt}) but we canceled it as we had also triggered SOAR on a brighter target the same night, which was eventually observed. Additionally, we first triggered SOAR observation 4116376 using the 400M1 grating of the Goodman Spectrograph, but decided to cancel it and trigger observation 4116378 with the 400M2 grating to benefit from the longer wavelength coverage provided by the latter mode. The data, including automatically reduced products for SOAR and Las Cumbres, were available as soon as the observations completed (within 1-2 hr of the triggers for spectra and even sooner for imaging). These were downloaded automatically into GOATS{\footnote{Currently, data retrieval for Blanco is not yet supported in GOATS.}}, and for Gemini data, we reduced them using the integrated DRAGONS software. 

At the time of this end-to-end campaign, Blanco had been integrated into AEON with its NEWFIRM instrument but not yet with DECam and the latter was mounted during those two nights. The workflow for Blanco was therefore a little different. Template observing scripts for DECam were prepared in advance and once the NEWFIRM trigger was received at Blanco, the Observatory staff helped execute the prepared observations after filling in the coordinates of the target. The follow-up data reduced automatically by the Community Pipeline \citep{community_pipeline} are available from the NOIRLab Astro Data Archive. The latter is not yet integrated into GOATS. Note, however, that starting semester 2026B, DECam on Blanco will also be available in AEON mode.

In total, we followed up on 18 alerts from 16 unique objects and obtained spectra for 7 targets, while the rest were imaging follow-up. All the follow-up data, including the automatically reduced data products are public; the program IDs are G-2025B-0529-D (Gemini), 2025B-037 (SOAR), 2026A-942051 (Blanco), and DDT2025B-006 (Las Cumbres). Details of the follow-up data are given in Table~\ref{tab:e2e}. We discuss our classification results below for targets with spectra from NOIRLab facilities. These spectra will be uploaded to TNS.

\paragraph{LSST-AP-DO-314051320244339039 / 2026hnt}

\begin{figure*}
\gridline{\fig{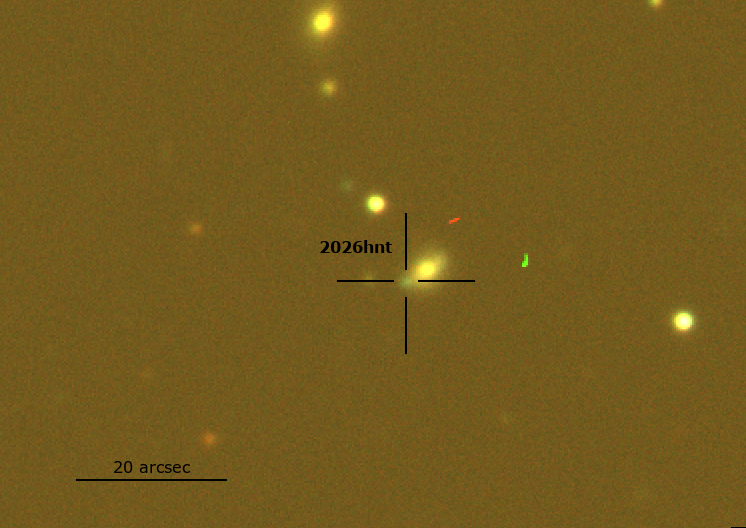}{0.45\textwidth}{(a)}
\fig{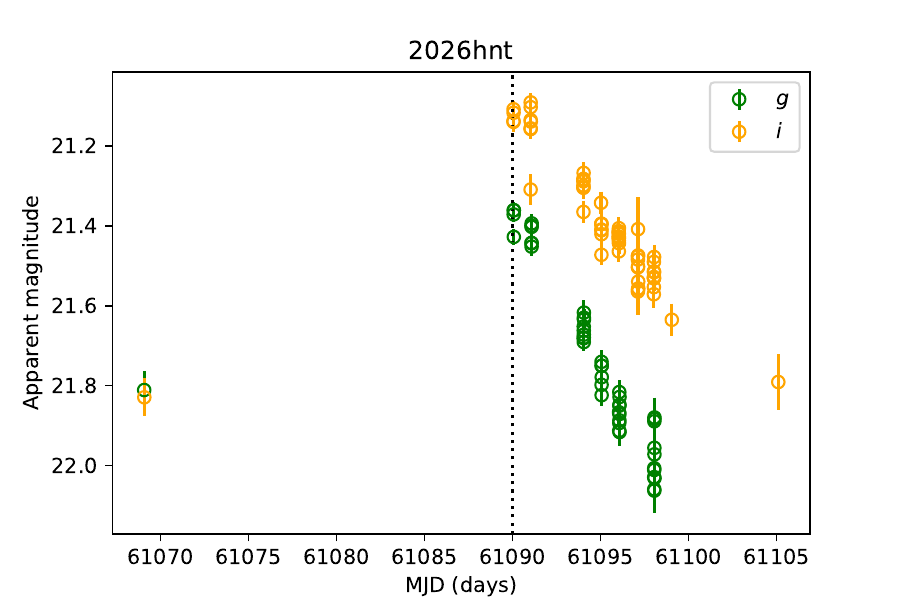}{0.55\textwidth}{(b)}
}
\gridline{\fig{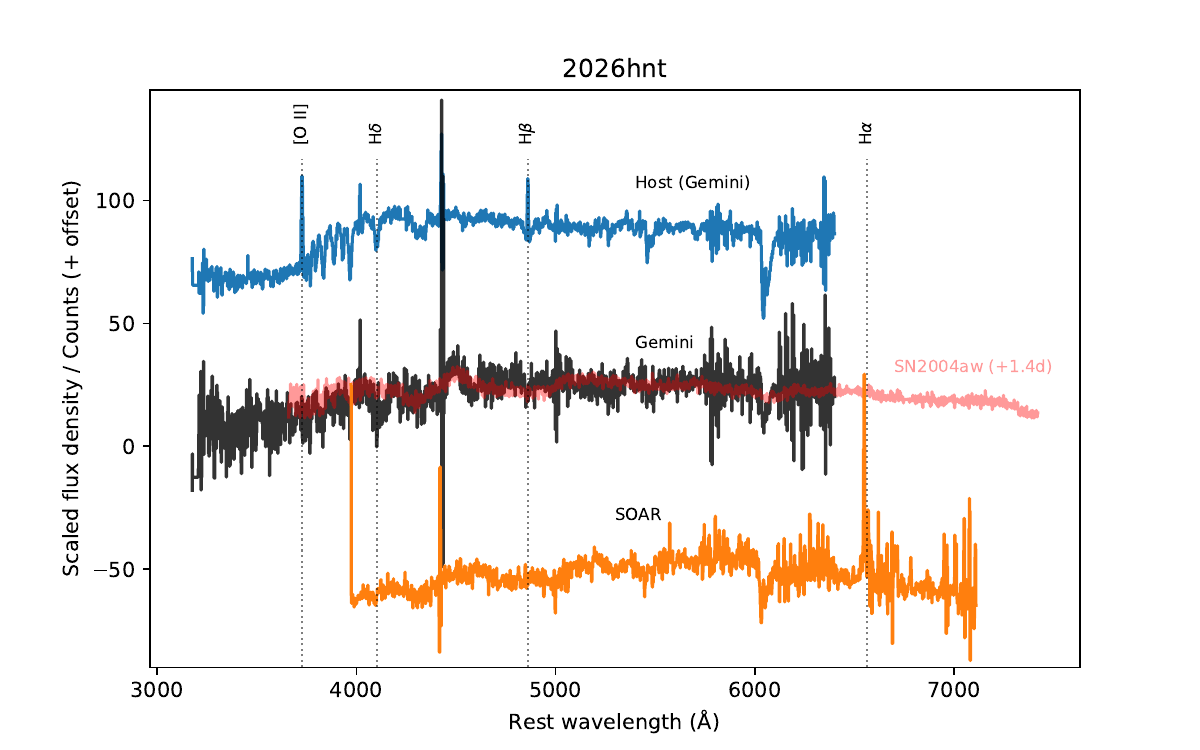}{\textwidth}{(c)}}
\caption{Follow-up data products for 2026nht. ({\it a}) Composite image showing 2026hnt and its host galaxy, created using its DECam follow-up data taken on 2026-02-19 UT (north is up and east is to the left). ({\it b}) Rubin/LSST light curve of 2026hnt; the vertical dotted line marks the first night of NOIRLab's end-to-end run when follow-up data were gathered using GOATS. ({\it c}) Follow-up spectra of 2026hnt obtained with various facilities of NOIRLab on 2026-02-19 UT. The spectrum of the host galaxy extracted interactively from the Gemini follow-up data using DRAGONS is also shown. The best-matching template spectrum of a supernova Type Ic (SN2004aw at 1.4~d from peak; \citealt{Modjaz-2014}) based on the  results from GELATO using the Gemini spectrum of 2026hnt is highlighted in light red. Note that the SOAR Goodman pipeline \citep{Torres} does not perform flux calibration, so the SOAR spectrum shown here, which has been automatically reduced by this pipeline, is not flux calibrated as compared to the other spectra.}    
\label{fig:2026hnt_data}
\end{figure*}

Fig.~\ref{fig:2026hnt_data} shows the follow-up images and spectra of this transient obtained on the first night of the end-to-end campaign. We reported its discovery to TNS using GOATS \citep{e2e_2026hnt} and it has been assigned the TNS name AT~2026hnt. The LCO follow-up spectrum, automatically reduced by the BANZAI pipeline \citep{McCully}, has very low S/N, so we do not use it for classification. The SOAR spectrum automatically reduced by the Goodman pipeline \citep{Torres}, is contaminated by light from the host galaxy, as indicated by the presence of the narrow {H}$\alpha$+[N II] lines. The wavelength coverage of the Gemini spectra did not extend to this region; nevertheless, with the interactive DRAGONS reduction in GOATS, we isolated the host and transient spectra, which are shown in the plot. The host spectrum shows strong Balmer absorption as well as [O II] emission simultaneously and matches an e(a)-type dusty starburst galaxy \citep{Poggianti-2009}; we obtain equivalent width of $\approx 11$~\AA\ and 7~\AA\ for [O II] and H$\delta$, respectively, and measure a redshift of $\approx 0.259$ that appears to be consistent with SOAR. We use GELATO{\footnote{\url{gelato.tng.iac.es}}} \citep{Harutyunyan-2008} to classify the transient spectrum from Gemini and obtain a match with supernova Type Ic SN2004aw \citep{Modjaz-2014}.   

\paragraph{LSST-AP-DO-313998539799134474 / 2026hob}

\begin{figure*}
\gridline{\fig{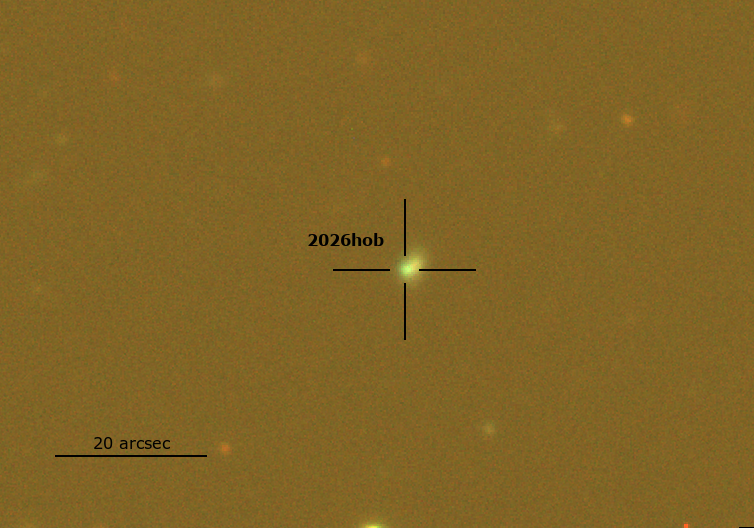}{0.47\textwidth}{(a)}
    \fig{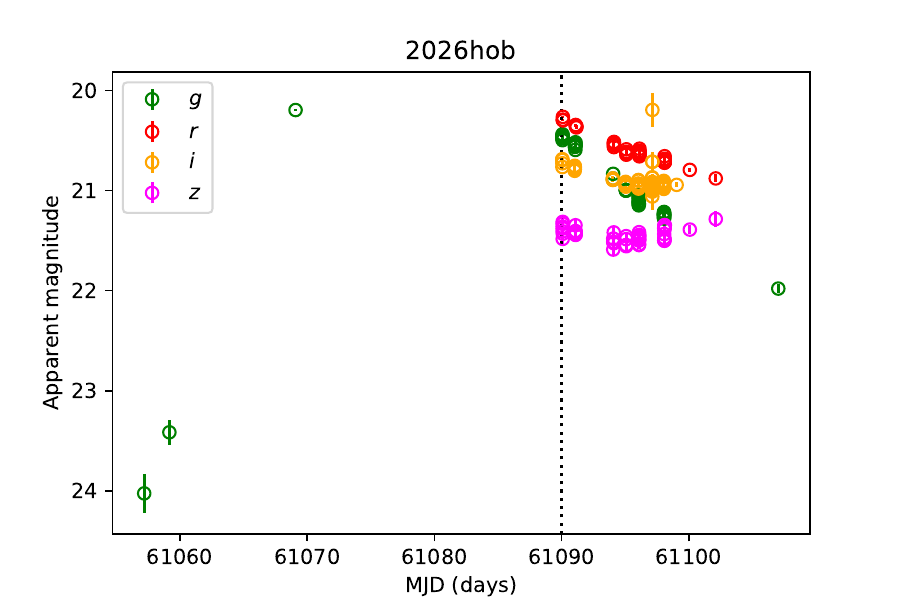}{0.55\textwidth}{(b)}}
\gridline{\fig{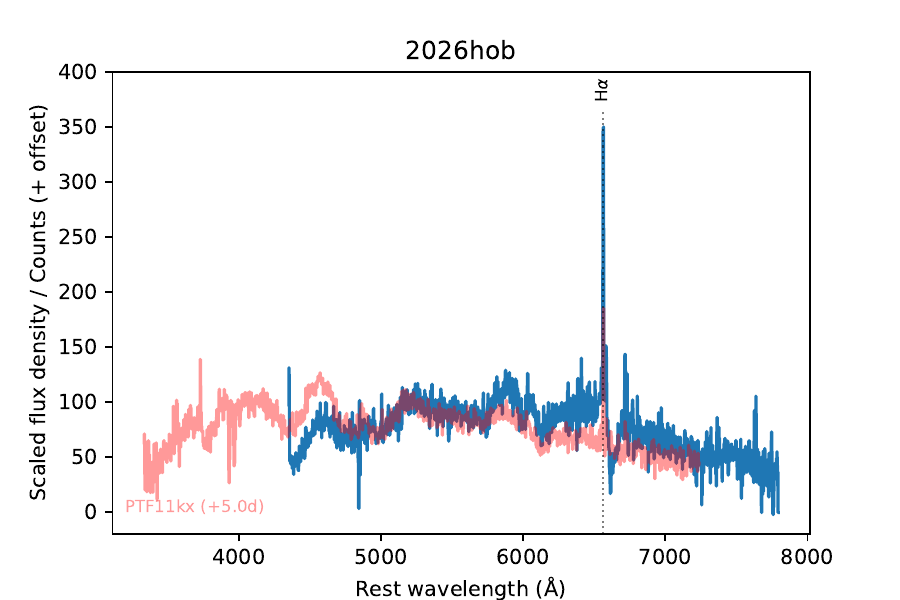}{0.6\textwidth}{(c)}}
\caption{({\it a}) Composite image showing 2026hob and its host galaxy, created using the DECam follow-up data taken on 2026-02-19 UT (north is up and east is to the left). 
({\it b}) Rubin/LSST light curve of 2026hob; the vertical dotted line marks the epoch (2026-02-19 UT) when follow-up data were gathered using GOATS as part of the NOIRLab end-to-end campaign. ({\it c}) Follow-up spectrum of 2026hob obtained with SOAR is shown in blue and the best-matching template spectrum (peculiar supernova Type Ia PTF11kx at 5~d from peak; \citealt{Dilday-2012}) from GELATO is highlighted in light red.}  
\label{fig:2026hob_data}
\end{figure*}

Fig.~\ref{fig:2026hob_data} shows the SOAR follow-up spectrum (automatically reduced by the Goodman pipeline) of this transient obtained on the first night of the end-to-end campaign. Its discovery was also reported to TNS by us using GOATS \citep{e2e_2026hnt} and it has been assigned the TNS name AT~2026hob. Narrow H$\alpha$ line from the host galaxy can be clearly seen in the spectrum, from which we measure a redshift of 0.147. This spectrum has been classified by GELATO as a supernova Type Ib/c based on the goodness-of-fit-weighted relative frequencies of the 20 best-fitting templates, but the result is uncertain. Nevertheless, among these templates, the peculiar supernova Type Ia PTF11kx provides a particularly good match, as shown in Fig.~\ref{fig:2026hob_data}. In fact, the light curve also appears to show a secondary peak in the $i$ and $z$ bands (panel (a) of Fig.~\ref{fig:2026hob_data}), characteristic of supernovae Type Ia,  which supports this classification. Note that although the SOAR spectrum is not flux-calibrated in the Goodman Pipeline reduction, this does not affect the template matching performed by GELATO (see \citealt{Harutyunyan-2008} for details).

\paragraph{LSST-AP-DO-170019716255973493 / 2026eio}
\begin{figure*}
\gridline{\fig{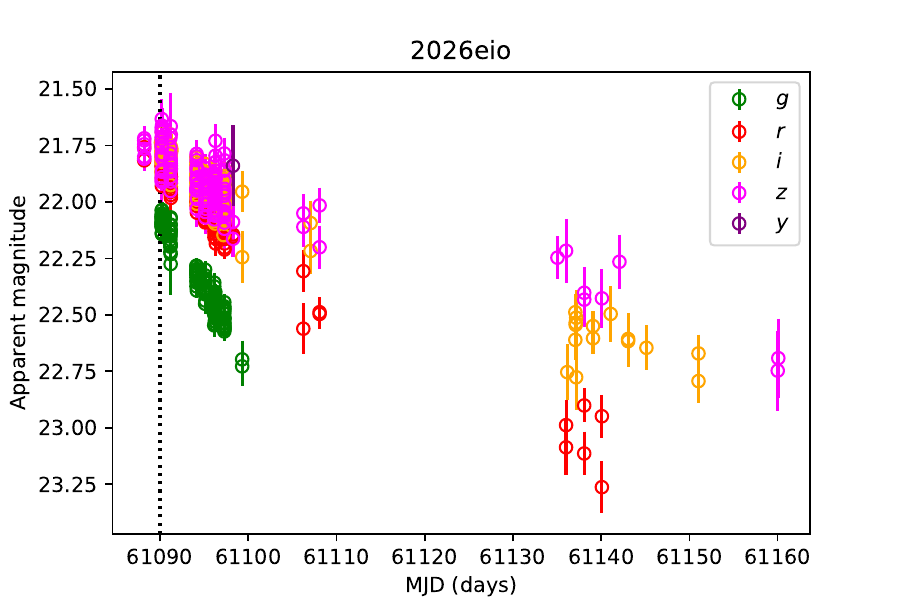}{0.6\textwidth}{(a)}}
\gridline{\fig{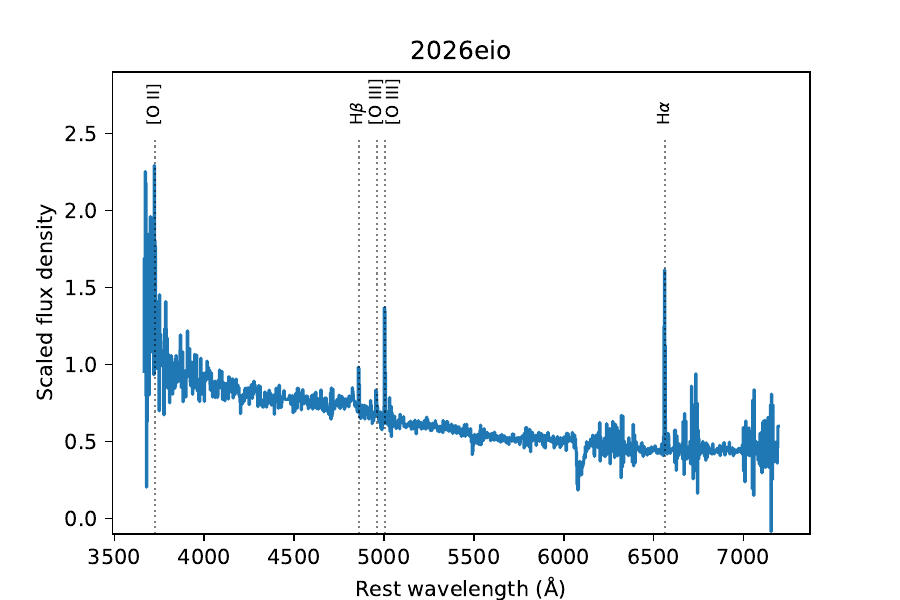}{0.6\textwidth}{(b)}
    \raisebox{0.12\height}{\fig{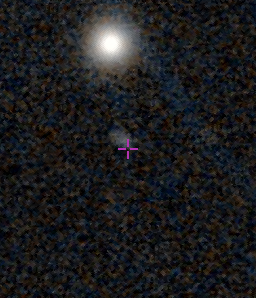}{0.26\textwidth}{(c)}}
}
\caption{({\it a}) Rubin/LSST light curve of 2026eio; the vertical dotted line marks the epoch (2026-02-19 UT) when follow-up data were gathered using GOATS. ({\it b}) Follow-up spectrum of 2026eio obtained with Gemini. Although follow-up imaging data were obtained with Las Cumbres 1-m Sinistro for this target, it was not detected in the images as they were too shallow. ({\it c}) Archival PanSTARRS color image cutout ($30''$ field-of-view) showing the location of the transient.}    
\label{fig:2026eio_data}
\end{figure*}

We selected and followed up this transient on the first night of the NOIRLab end-to-end campaign, i.e., 2026-02-19 UT. Fig.~\ref{fig:2026eio_data} shows its Gemini follow-up spectrum, reduced interactively using the integrated DRAGONS in GOATS. The discovery of this transient was later reported to TNS on 2026-02-24 UT by \citet{Murphey} based on Young Supernova Experiment's DECam survey data, and it has been assigned the TNS name AT~2026eio. For our Gemini follow-up, we oriented the slit to minimize contamination from the host galaxy, yet narrow emission lines from the host, such as [O II], [O III], H$\alpha$, and H$\beta$, are clearly visible in the spectrum. This indicates recent star formation at the location of the transient, which may suggest a massive-star origin. Using the host H$\alpha$ line, we measure a redshift of $\approx 0.251$. No apparent lines from the transient are, however, visible, preventing its classification. The spectrum has been flux calibrated during DRAGONS reduction. Given the shape of this spectrum, the transient likely exhibited an intrinsically featureless blue continuum. Since it is significantly offset from the nucleus of the host galaxy (see panel (c) of Fig.~\ref{fig:2026eio_data}), a tidal disruption event is unlikely. Furthermore, the spectrum was obtained around peak light, when the transient reached an absolute magnitude of $M_{z}\sim-18.8$~mag (not corrected for extinction). The light-curve decline rate is around 0.06~mag/d, which is faster than that of hydrogen-rich core-collapse supernovae \citep[e.g.,][]{Anderson-2014}, but appears within the range of stripped-envelope supernovae \citep{Taddia-2018}, some of which show blue featureless spectra at peak such as iPTF16asu \citep{Whitesides-2017}.   

\paragraph{LSST-AP-DO-314051320305156133 / 2026dmz}
\begin{figure*}
\gridline{\fig{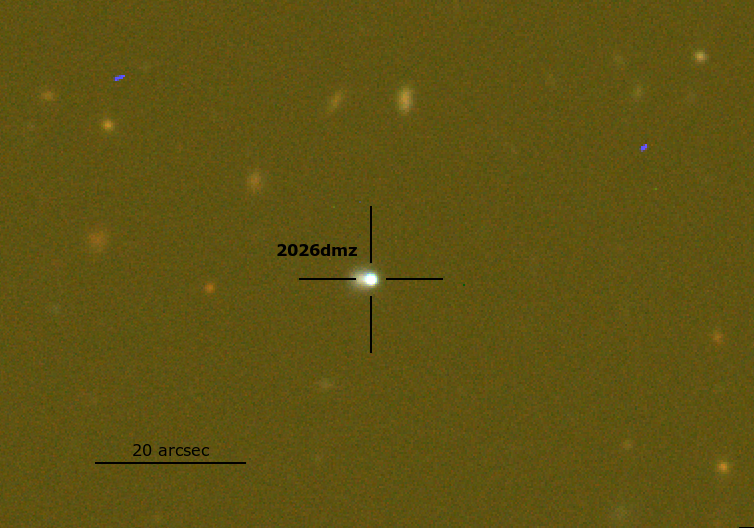}{0.47\textwidth}{(a)}
    \fig{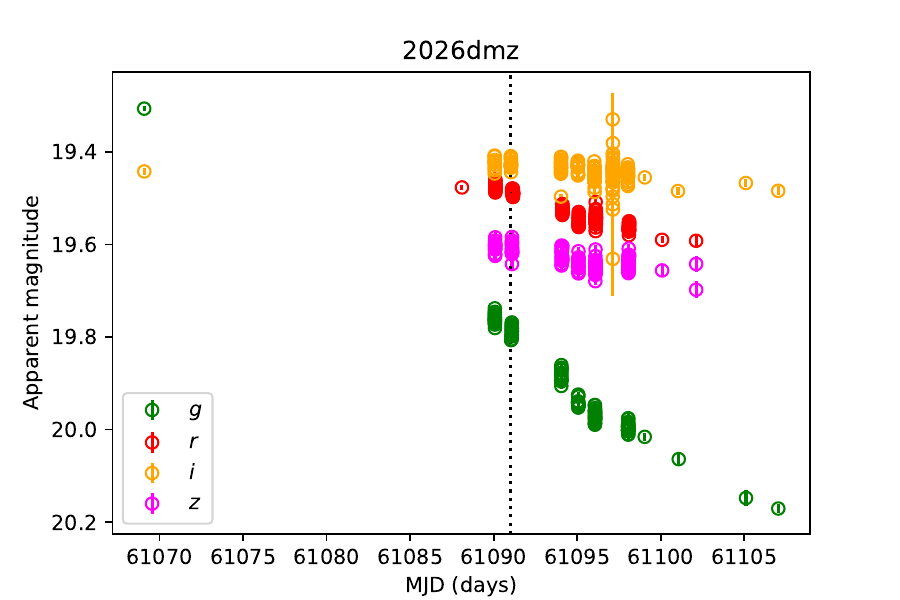}{0.55\textwidth}{(b)}}
\gridline{\fig{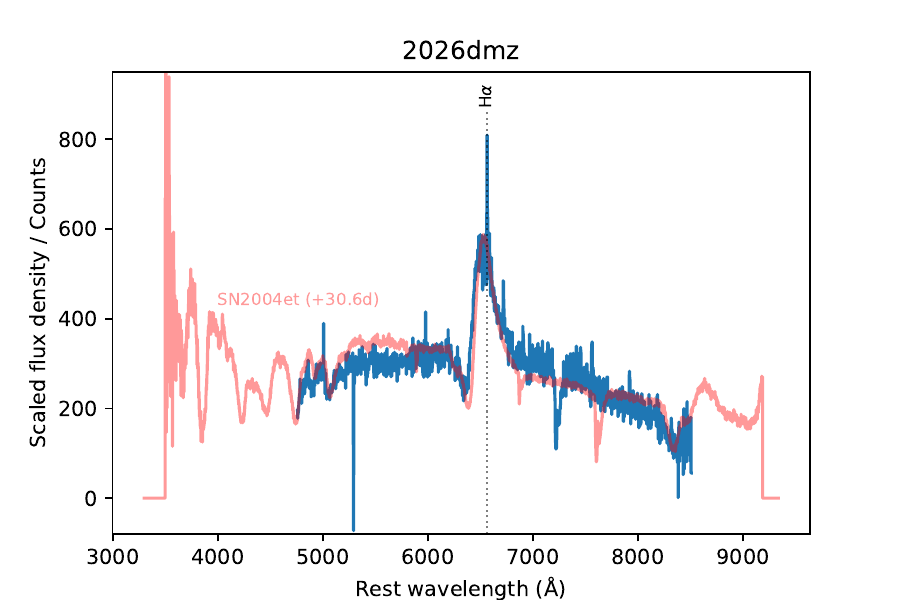}{0.6\textwidth}{(c)}}
\caption{({\it a}) Composite image showing 2026dmz and its host galaxy, created using the DECam follow-up data taken on 2026-02-20 UT (north is up and east is to the left). 
({\it b}) Rubin/LSST light curve of 2026dmz; the vertical dotted line marks the epoch (2026-02-20 UT) when follow-up data were gathered using GOATS. ({\it c}) Follow-up spectrum of 2026dmz obtained with SOAR (blue) and the best-matching template spectrum (supernova Type II SN2004et at 30.6~d post explosion; \citealt{Sahu-2006}) from GELATO is highlighted in light red.}
\label{fig:2026dmz_data}
\end{figure*}

Fig.~\ref{fig:2026dmz_data} shows the SOAR follow-up spectrum (automatically reduced by the Goodman pipeline) of this transient obtained on the second night of the end-to-end campaign. Its discovery was reported to TNS on 2026-02-18 UT by \citet{Tonry} based on detection by the ATLAS survey, and it has been assigned the TNS name AT~2026dmz. We selected it independently for follow-up when it was flagged by ANTARES based exclusively on Rubin/LSST alert stream during the end-to-end campaign. Narrow H$\alpha$ line from the host galaxy is clearly visible in the spectrum, from which we measure a redshift of 0.053. We classify it as supernova Type IIP using GELATO. The best-matching template spectrum from GELATO, that of SN2004et \citep{Sahu-2006}, is also shown in the figure.

\paragraph{LSST-AP-DO-170019716341956650 / 2026eil}
\begin{figure*}
\gridline{\fig{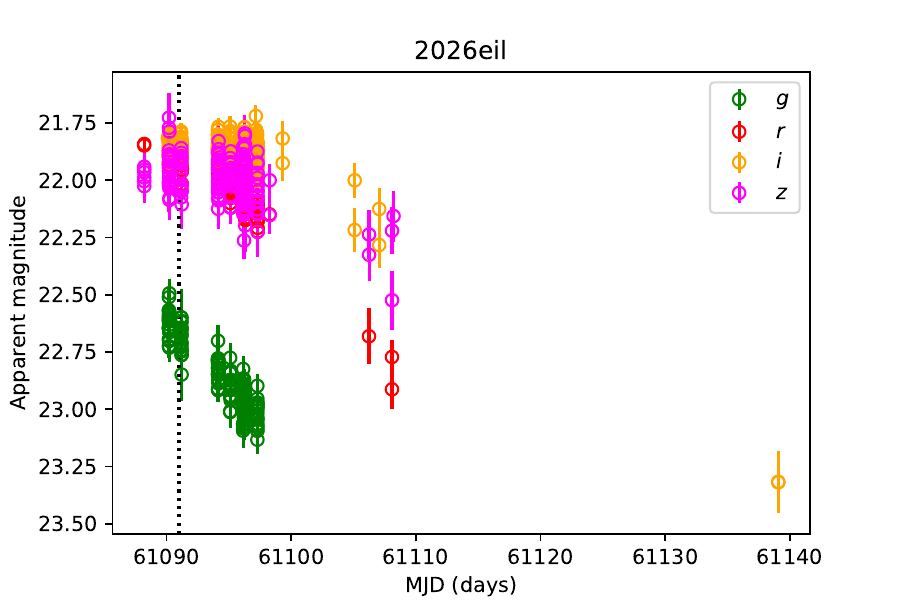}{0.6\textwidth}{(a)}}
\gridline{\fig{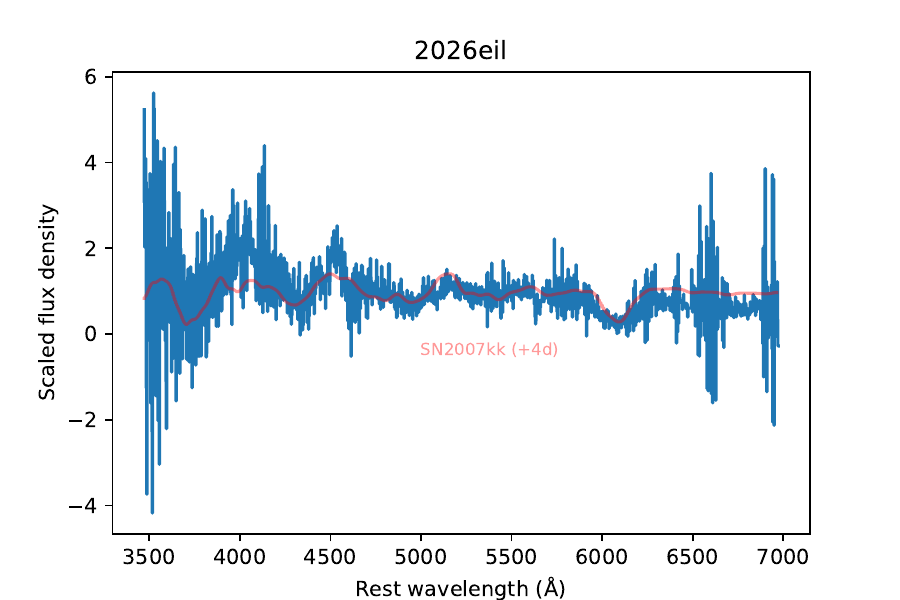}{0.6\textwidth}{(b)}
    \raisebox{0.12\height}{\fig{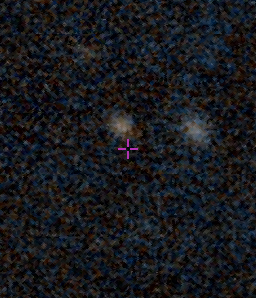}{0.26\textwidth}{(c)}}
}
\caption{({\it a}) Rubin/LSST light curve of 2026eil; the vertical dotted line marks the epoch (2026-02-20 UT) when follow-up data were gathered using GOATS. ({\it b}) Follow-up spectrum of 2026eil obtained with Gemini (blue), overplotted with the best-matching template of supernova Type Ia SN2007kk from SNID (red). No follow-up imaging data was obtained for this target.
({\it c}) Archival PanSTARRS color image cutout ($30''$ field-of-view) showing the location of the transient.
}    
\label{fig:2026eil_data}
\end{figure*}

We selected and followed up this transient on the second night of the end-to-end campaign (2026-02-20 UT). Fig.~\ref{fig:2026eil_data} shows our Gemini follow-up spectrum, reduced interactively using the integrated DRAGONS in GOATS. The discovery of this object was later reported to TNS on 2026-02-24 UT by \citet{Murphey} based on Young Supernova Experiment's DECam survey data, and has been assigned the TNS name AT~2026eil. We classify our follow-up spectrum using SNID \citep{Blondin} and obtain a result of supernova Type Ia at a redshift of 0.35 and phase +4~d from peak. 

\paragraph{LSST-AP-DO-170019717277810735 / 2026ctw}
\begin{figure*}
\gridline{\fig{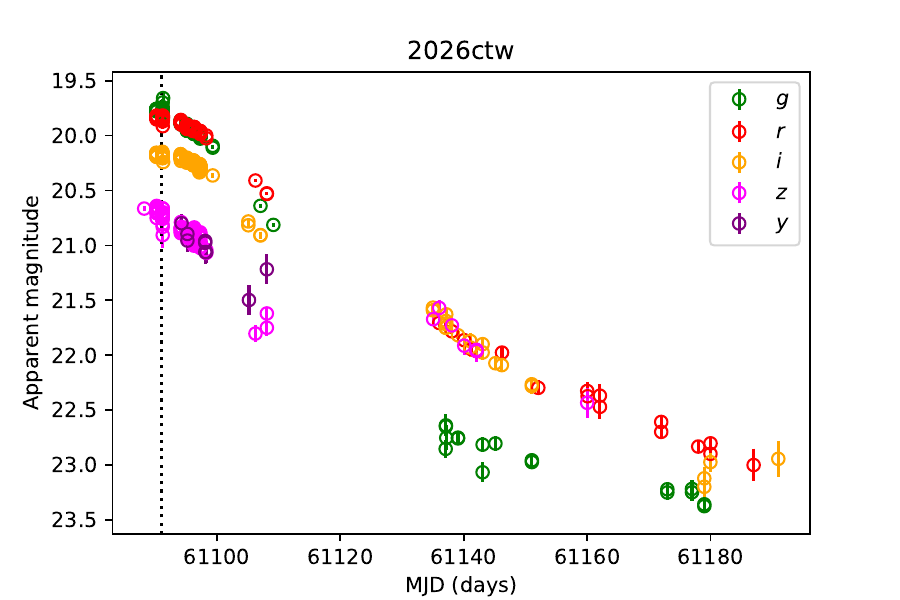}{0.6\textwidth}{(a)}}
\gridline{\fig{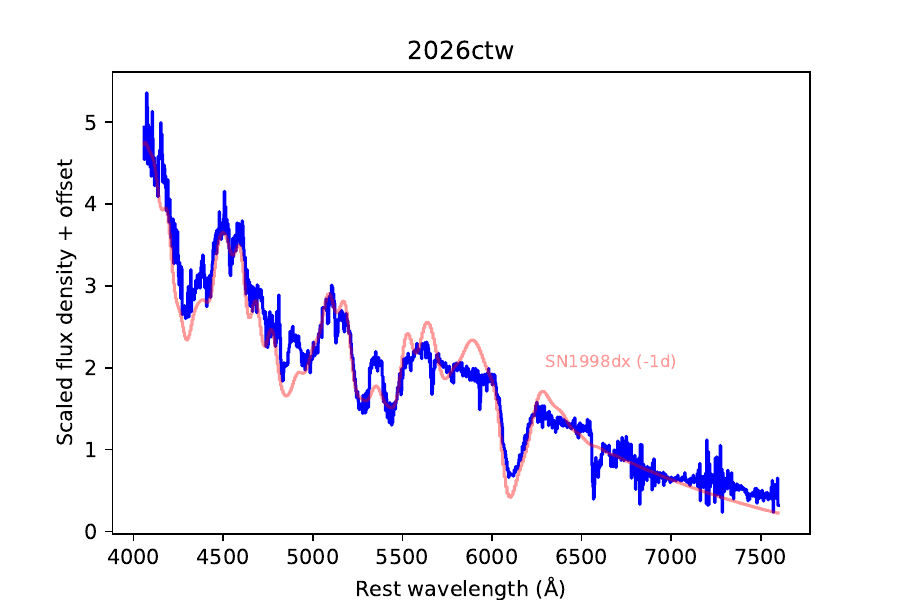}{0.6\textwidth}{(b)}
    \raisebox{0.12\height}{\fig{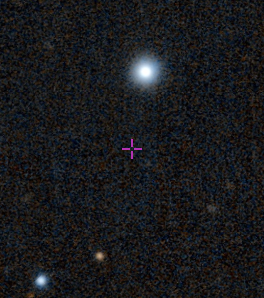}{0.27\textwidth}{(c)}}
}
\caption{({\it a}) Rubin/LSST light curve of 2026ctw; the vertical dotted line marks the epoch (2026-02-20 UT) when follow-up data were gathered using GOATS. ({\it b}) Follow-up spectrum of 2026ctw obtained with Gemini (blue) and the best-matching template of supernova Type Ia SN1998dx from SNID (red). No follow-up imaging data was obtained for this target.
({\it c}) Archival PanSTARRS color image cutout ($60''$ field-of-view) showing the location of the transient. There is no apparent host visible.
}    
\label{fig:2026ctw_data}
\end{figure*}

The discovery of this transient was reported to TNS on 2026-02-10 UT by \cite{Hall-2026} based on detection by DECam. We independently selected it for follow-up when it was flagged by ANTARES in the Rubin/LSST alert stream on the second night of the end-to-end campaign. Fig.~\ref{fig:2026ctw_data} shows our Gemini follow-up spectrum from 2026-02-20 UT, reduced interactively using the integrated DRAGONS in GOATS. We classify it with SNID \citep{Blondin} and obtain a result of supernova Type Ia at a redshift of 0.156 and phase -1~d from peak. Note that a similar classification has also been reported to TNS by \citet{Rose-2026}, based on a later spectrum taken on 2026-02-23 UT with the Palomar 200-inch Hale Telescope.

\vspace{1em}
For LSST-AP-DO-170028526873870371 / 2025ahkl, we only triggered LCO/FLOYDS on UT 2026-02-20. However, there was likely some issue with the target acquisition and the target was not on the slit, so we could not obtain its spectrum.

\section{Future direction for GOATS}\label{sec:future}

NOIRLab provides a suite of telescopes with apertures ranging from 3-m to 8-m, covering both hemispheres. These facilities, namely, Gemini, SOAR, Blanco, and WIYN, are invaluable for follow-up of MMA/TDA events by the broader astronomy community. In the following, we describe features that could potentially be implemented in GOATS in the future. 

GOATS can be extended to support end-to-end follow-up observations not only from Gemini, SOAR, and external observatories with available TOM-Toolkit plugins, but across all NOIRLab facilities. This will be facilitated by integrating Blanco (DECam), WIYN, and, in the future, Rubin and the US-ELT into AEON, enabling GOATS to transition into a \emph{General} Observation and Analysis of Targets System for NOIRLab.   

GOATS can greatly benefit from incorporating an artificial intelligence (AI) agent, which will optimize follow-up configurations across facilities available to the user. For example, given an unclassified  target, the agent can automatically generate {\it fully prepared observation} recommendations, selecting the appropriate facility/instrument and configuration (e.g., observing mode, observing filter, exposure time, grating, central wavelength, etc.\,) taking into account  real-time observing conditions and the range of plausible transient classifications. If the classification is known, the agent incorporates it along with the expected phase of the object, inferred directly from the light curve of the target in GOATS, to inform its recommendations. For instance, it may select a higher-resolution grating when it predicts the transient to be in the nebular phase, where spectral lines are much narrower than in the photospheric phase. Furthermore, the AI can provide even more advanced  assistance by orchestrating the entire workflow, coordinating observations across multiple facilities to both classify the transient and monitor its evolution.

For the added facilities, besides the triggering capability, GOATS can also integrate services for data retrieval (in particular, the NOIRLab Astro Data Archive), reduction, and analysis (for example, the Rubin Science Platform), thereby fully closing the follow-up loop. Even in the absence of a unified archive for the various telescope facilities, this tool can ensure that users are able to find and access their follow-up data in a uniform manner. 

Furthermore, for MMA/TDA studies, a user often needs to perform joint analysis of follow-up data collected from multiple facilities. A lack of a common data reduction framework or pipeline presents a significant bottleneck. Developing a standardized framework for data reduction will greatly improve user efficiency and facilitate scientific reproducibility. DRAGONS provides a suitable solution, as it is general enough to support not only Gemini data but also data from other telescopes via third-party extensions. This software can thus be adopted for reduction of data from the rest of NOIRLab facilities. Our integration of DRAGONS into GOATS will thus prove to be a powerful tool to maximize the scientific return from all NOIRLab facilities. 

Additionally, GOATS can be made interoperable with external MMA services. These include communication services like Hermes{\footnote{\url{https://hermes.lco.global}}}, developed by Las Cumbres in collaboration with the Scalable CyberInfrastructure for Multi-Messenger Astronomy (SCiMMA; \citealt{SCiMMA}), which allows dissemination of MMA/TDA event information and sharing of targets and follow-up data, and the Gravitational Wave Treasure Map \citep{TMap}, which provides a messaging service for reporting and coordinating planned follow-up observations to search for electromagnetic counterparts of gravitational wave events. In particular, via Hermes, we can enable sharing of target and follow-up data between different users running GOATS locally on their own systems{\footnote{For example, user A running a local instance of GOATS (i.e., on their own localhost) can share target information and data, including proprietary data, with another user B, who is running a separate GOATS instance on their own localhost.}} and also harvesting of {\it non-localized events}, such as localization maps associated with gravitational wave and neutrino alerts.

While the functionality for performing spectral data analysis on the frontend is part of the current GOATS development effort, imaging data analysis is not yet supported. This capability could be added by implementing interactive photometry tools. 
Specifically, the existing visualization components in GOATS can be extended so that users may display and inspect follow-up images, mark sources of interest, and perform forced photometry directly within the interface. Additionally, GOATS could provide a one-click option to extract forced-photometry measurements derived from all relevant observations available within Astro Data Archive, including from automatically generated DECam difference images \citep{Fu-2024}.

Finally, the present GOATS system is primarily designed for local hosting, which enhances security and performance by eliminating dependence on external servers and allows users to maintain better control over their data. Nevertheless, it already provides multi-user support and we can deploy it online, thereby offering scalability and accessibility for more extensive operations. This in particular would facilitate community coordination.

\section{Summary}\label{sec:summary}
Follow up of MMA/TDA events requires responsive observational capabilities coupled with equally responsive and user-friendly software tools. As part of AEON, which comprises telescopes capable of rapid response to such events, the twin Gemini telescopes are the only 8-m class facility especially enabled for agile transient follow-up, and also available to the whole astronomy community. In this paper, we describe GOATS, which is an end-to-end software tool we have built for planning, triggering, reducing and analyzing follow-up observations with AEON facilities, in particular Gemini, thereby lowering the entry barrier for users. GOATS provides a seamless connection between observing facilities, data, and investigators, thus enabling the MMA/TDA community to make  discoveries from the gamut of revolutionary time-domain and gravitational wave surveys.  

\begin{longrotatetable}
\begin{deluxetable*}{llllclll}
\tablecaption{Summary of Rubin/LSST alert follow-up observations from the NOIRLab-Las Cumbres end-to-end campaign \label{tab:e2e}}
\tablehead{
\colhead{RA (J2000)} & \colhead{DEC (J2000)} & \colhead{Obs. UT date} & \colhead{Facility} & \colhead{Instrument} & \colhead{\parbox{1cm}{Grating/\\Passband}} & \colhead{Obs. ID} & \colhead{ANTARES ID with link}
}
\startdata
\multicolumn{8}{l}{Target LSST-AP-DO-314051320244339039/2026hnt}\\
 03:55:57.56    &   -50:50:35.59 & 2026-02-19 & Gemini South & GMOS & B480 & G-2025B-0529-D-0223 & \href{https://antares.noirlab.edu/loci/ANT20265qo8lfenipdc}{ANT20265qo8lfenipdc}\\ 
  & & 2026-02-19 & Las Cumbres & 1-m Sinistro & $gri$ & DDT2025B-006 & \\ 
  & & 2026-02-19 & Las Cumbres & 2-m FLOYDS & XD & DDT2025B-006 & \\
  & & 2026-02-19 & SOAR & Goodman Spectrograph & 400M2 & SOAR2025B-037 & \\
  & & 2026-02-19 & Blanco & DECam & $griz$ & 2026A-942051 & \\
  & & 2026-02-20 & Blanco & DECam & $griz$ & 2026A-942051 & \\
  \hline
\multicolumn{8}{l}{Target LSST-AP-DO-313998539799134474/2026hob}\\
04:12:03.32 &   -49:28:15.46 & 2026-02-19 & SOAR & Goodman Spectrograph & 400M2 & SOAR2025B-037 & \href{https://antares.noirlab.edu/loci/ANT2026hrd3s53ks4jt}{ANT2026hrd3s53ks4jt}\\ 
 & & 2026-02-19 & Blanco & DECam & $griz$ & 2026A-942051 & \\
 \hline
\multicolumn{8}{l}{Target LSST-AP-DO-170019716255973493/2026eio}\\
09:57:21.06 &   00:57:37.53 & 2026-02-19 & Gemini North & GMOS & R400 & G-2025B-0529-D-0240 & \href{https://antares.noirlab.edu/loci/ANT2026rim8k1jgs1mg}{ANT2026rim8k1jgs1mg}\\
 & & 2026-02-19 & Las Cumbres & 1-m Sinistro & $gi$ & DDT2025B-006 & \\ 
 \hline
\multicolumn{8}{l}{Target LSST-AP-DO-313963359491850270/2026ejv}\\ 
04:12:42.07 &   -47:23:31.47 & 2026-02-19 & Blanco & DECam & $griz$ & 2026A-942051 & \href{https://antares.noirlab.edu/loci/ANT2026otihcvulecis}{ANT2026otihcvulecis}\\
 \hline
\multicolumn{8}{l}{Target LSST-AP-DO-313972153257558986/2026hoj}\\
04:04:16.34 &   -50:11:44.43 & 2026-02-19 & Blanco & DECam & $griz$ & 2026A-942051 & \href{https://antares.noirlab.edu/loci/ANT2026ekd9ai4s2mz0}{ANT2026ekd9ai4s2mz0}\\
 \hline
\multicolumn{8}{l}{Target LSST-AP-DO-170019696272736349/2026fku}\\
04:05:59.31 &   -49:14:44.84 & 2026-02-19 & Blanco & DECam & $griz$ & 2026A-942051 & \href{https://antares.noirlab.edu/loci/ANT20263f6uz7n8dsto}{ANT20263f6uz7n8dsto}\\
 & & 2026-02-20 & Blanco & DECam & $griz$ & 2026A-942051 & \\
 \hline
\multicolumn{8}{l}{Target LSST-AP-DO-313967752817672227/2026emp}\\
04:14:5.76  &   -48:05:02.15 & 2026-02-19 & Blanco & DECam & $griz$ & 2026A-942051 & \href{https://antares.noirlab.edu/loci/ANT2026bay53hmb6fsn}{ANT2026bay53hmb6fsn}\\
 \hline
\multicolumn{8}{l}{Target LSST-AP-DO-314051320305156133/2026dmz}\\
04:01:30.64 &   -48:49:19.18 & 2026-02-20 & SOAR & Goodman Spectrograph & 400M2 & SOAR2025B-037 & \href{https://antares.noirlab.edu/loci/ANT2026wun1eleya11n}{ANT2026wun1eleya11n}\\
 & & 2026-02-20 & Las Cumbres & 1-m Sinistro & $gr$ & DDT2025B-006 & \\
 & & 2026-02-20 & Blanco & DECam & $griz$ & 2026A-942051 & \\
 \hline
\multicolumn{8}{l}{Target LSST-AP-DO-170019716341956650/2026eil}\\
10:03:21.88 &   02:42:5.79 & 2026-02-20 & Gemini South & GMOS & R400 & G-2025B-0529-D-0331 & \href{https://antares.noirlab.edu/loci/ANT2026bx5bf2bye1c7}{ANT2026bx5bf2bye1c7}\\
 \hline
\multicolumn{8}{l}{Target LSST-AP-DO-170019717277810735/2026ctw}\\
10:06:57.67 &   01:55:37.37 & 2026-02-20 & Gemini North & GMOS & R400 & G-2025B-0529-D-0332 & \href{https://antares.noirlab.edu/loci/ANT2026km1wlq37d7g5}{ANT2026km1wlq37d7g5}\\
 \hline
\multicolumn{8}{l}{Target LSST-AP-DO-170028526873870371/2025ahkl}\\
12:27:13.98 &   08:28:56.70 & 2026-02-20 & Las Cumbres & 2-m FLOYDS & XD & DDT2025B-006 & \href{https://antares.noirlab.edu/loci/ANT2026m55ceon1vrn6}{ANT2026m55ceon1vrn6}\\
 \hline
\multicolumn{8}{l}{Target LSST-AP-DO-314051320414732370/2026hol}\\
04:15:26.44 &   -48:10:23.79 & 2026-02-20 & Blanco & DECam & $griz$ & 2026A-942051 & \href{https://antares.noirlab.edu/loci/ANT2026jes2tm1rkkfw}{ANT2026jes2tm1rkkfw}\\
 \hline
\multicolumn{8}{l}{Target LSST-AP-DO-314051321263554927/2026hom}\\
04:13:28.04 &   -45:53:55.46 & 2026-02-20 & Blanco & DECam & $griz$ & 2026A-942051 & \href{https://antares.noirlab.edu/loci/ANT2026nytdm6yio0il}{ANT2026nytdm6yio0il}\\
 \hline
\multicolumn{8}{l}{Target LSST-AP-DO-170019696299475118/2026fgr}\\
04:00:36.14 &   -48:25:10.72 & 2026-02-20 & Blanco & DECam & $griz$ & 2026A-942051 & \href{https://antares.noirlab.edu/loci/ANT2026lhc4eysvjz9o}{ANT2026lhc4eysvjz9o}\\
 \hline
\multicolumn{8}{l}{Target LSST-AP-DO-170028526654193673/2026hon}\\
12:29:42.62 &   07:06:8.17 & 2026-02-20 & Las Cumbres & 1-m Sinistro & $r$ & DDT2025B-006 & \href{https://antares.noirlab.edu/loci/ANT20263mmimjt7ncn9}{ANT20263mmimjt7ncn9}\\
 \hline
\multicolumn{8}{l}{Target LSST-AP-DO-170028526448672839/2026hoo}\\
12:21:11.37 &   07:49:31.57 & 2026-02-20 & Las Cumbres & 1-m Sinistro & $i$ & DDT2025B-006 & \href{https://antares.noirlab.edu/loci/ANT2026b6zv83zjo2rq}{ANT2026b6zv83zjo2rq}\\
\enddata
\label{tab:e2e}
\end{deluxetable*}
\end{longrotatetable}

\appendix
\section{Contributions to upstream external software projects}\label{sec:appendix}
We have been an active upstream contributor to the TOM-Toolkit base library. We submitted several pull requests addressing bugs, usability gap, and architectural limitations encountered during GOATS development, which were merged into the base library by the TOM-Toolkit developers. 

Our contributions have addressed longstanding issues, such as in data product deletion that caused orphaned files and filename collisions, a formatting typo that had produced incorrect timestamps across previous TOM-Toolkit deployments, and a bug that prevented comment deletion. Furthermore, many of our contributions led to usability improvement in the base library, for example, the ability to navigate to specific tabs in the target detail view, batch selection of targets from alert broker queries, configurable pagination display, customization of visualization theme, and asynchronous loading of the facility status page to make it more responsive. We also contributed pull requests addressing deeper issues in the base library's architecture. These include fixing a bug in API route registration that crashed applications using custom endpoints, and extending the data processing pipeline to support runtime selection of processing strategies.

We also co-maintain the conda packages for \texttt{tomtoolkit} and \texttt{tom-tns}. Additionally, our extension of the \texttt{gpp-client} will enable users of existing TOM systems to add support for triggering Gemini observations through GPP.

\begin{acknowledgments}
This material is based upon work supported by the National Science Foundation through the Windows on the Universe (WoU-MMA) program.

The international Gemini Observatory, a program of NSF NOIRLab, is managed by the Association of Universities for Research in Astronomy (AURA) under a cooperative agreement with the U.S. National Science Foundation on behalf of the Gemini partnership: the U.S. National Science Foundation (United States), National Research Council (Canada), Agencia Nacional de Investigaci\'{o}n y Desarrollo (Chile), Ministerio de Ciencia, Tecnolog\'{i}a e Innovaci\'{o}n (Argentina), Minist\'{e}rio da Ci\^{e}ncia, Tecnologia, Inova\c{c}\~{o}es e Comunica\c{c}\~{o}es (Brazil), and Korea Astronomy and Space Science Institute (Republic of Korea). 

This material is based upon work supported in part by the National Science Foundation through Cooperative Agreements AST-1258333 and AST-2241526 and Cooperative Support Agreements AST-1202910 and 2211468 managed by the Association of Universities for Research in Astronomy (AURA), and the Department of Energy under Contract No. DE-AC02-76SF00515 with the SLAC National Accelerator Laboratory managed by Stanford University. Additional Rubin Observatory funding comes from private donations, grants to universities, and in-kind support from LSST-DA Institutional Members.

Based on observations obtained through the Astronomical Event Observatory Network (AEON), a joint endeavor of the Las Cumbres Observatory and of NSF NOIRLab, which is managed by the Association of Universities for Research in Astronomy (AURA) under a cooperative agreement with the U.S. National Science Foundation. 

Based in part on observations obtained at the Southern Astrophysical Research (SOAR) telescope, which is a joint project of the Minist\'{e}rio da Ci\^{e}ncia, Tecnologia e Inova\c{c}\~{o}es (MCTI/LNA) do Brasil, the US National Science Foundation’s NOIRLab, the University of North Carolina at Chapel Hill (UNC), and Michigan State University (MSU).

This project used data obtained with the Dark Energy Camera (DECam), which was constructed by the Dark Energy Survey (DES) collaboration. Funding for the DES Projects has been provided by the US Department of Energy, the U.S. National Science Foundation, the Ministry of Science and Education of Spain, the Science and Technology Facilities Council of the United Kingdom, the Higher Education Funding Council for England, the National Center for Supercomputing Applications at the University of Illinois at Urbana-Champaign, the Kavli Institute for Cosmological Physics at the University of Chicago, Center for Cosmology and Astro-Particle Physics at the Ohio State University, the Mitchell Institute for Fundamental Physics and Astronomy at Texas A\&M University, Financiadora de Estudos e Projetos, Fundação Carlos Chagas Filho de Amparo à Pesquisa do Estado do Rio de Janeiro, Conselho Nacional de Desenvolvimento Científico e Tecnológico and the Ministério da Ciência, Tecnologia e Inovação, the Deutsche Forschungsgemeinschaft and the Collaborating Institutions in the Dark Energy Survey.

The Collaborating Institutions are Argonne National Laboratory, the University of California at Santa Cruz, the University of Cambridge, Centro de Investigaciones Enérgeticas, Medioambientales y Tecnológicas–Madrid, the University of Chicago, University College London, the DES-Brazil Consortium, the University of Edinburgh, the Eidgenössische Technische Hochschule (ETH) Zürich, Fermi National Accelerator Laboratory, the University of Illinois at Urbana-Champaign, the Institut de Ciències de l’Espai (IEEC/CSIC), the Institut de Física d’Altes Energies, Lawrence Berkeley National Laboratory, the Ludwig-Maximilians Universität München and the associated Excellence Cluster Universe, the University of Michigan, NSF NOIRLab, the University of Nottingham, the Ohio State University, the OzDES Membership Consortium, the University of Pennsylvania, the University of Portsmouth, SLAC National Accelerator Laboratory, Stanford University, the University of Sussex, and Texas A\&M University.

Based on observations at NSF Cerro Tololo Inter-American Observatory, NSF NOIRLab, which is managed by the Association of Universities for Research in Astronomy (AURA) under a cooperative agreement with the U.S. National Science Foundation.

We are grateful to the TOM-Toolkit developers{\footnote{\url{https://lco.global/tomtoolkit/tom-team/}}}, especially Joey Chatelain and Lindy Lindstrom for their close collaboration with us throughout this project. 

We acknowledge the use of OpenAI ChatGPT and Atlassian Rovo to assist in improving the readability of the manuscript based on the original content of the authors; these tools were not used for any scientific analysis or citations.

\end{acknowledgments}

\facilities{Gemini, SOAR, Las Cumbres Observatory, CTIO:4m (Blanco), Astro Data Lab}
\software{\texttt{astropy} \citep{astropy-2013,  astropy-2018}, \texttt{matplotlib} \citep{matplotlib}, \texttt{specutils} \citep{specutils}, \texttt{jdaviz} \citep{jdaviz}}

\bibliography{references}{}
\bibliographystyle{aasjournalv7}

\end{document}